\documentclass[11pt]{article}
\pdfoutput=1 % if your are submitting a pdflatex (i.e. if you have
\usepackage{jheppub} % for details on the use of the package, please
\pdfoutput=1
\usepackage{epstopdf}
\usepackage{mathtools}
\usepackage{mathrsfs}
\usepackage{psfrag}
\usepackage{color}
\usepackage{hyperref}

\usepackage{slashed}
\usepackage{simplewick}
\usepackage{cancel}
\usepackage[utf8]{inputenc}
\usepackage{caption}

\usepackage{amsmath,bbm,array,amsfonts,graphicx,wrapfig,arydshln,lscape,float}

\usepackage{url}

\let\olditemize\itemize\renewcommand{\itemize}{\vspace{-2pt}\olditemize\setlength{\itemsep}{1pt}\setlength{\parskip}{0pt}\setlength{\parsep}{-0pt}}
\let\oldenumerate\enumerate\renewcommand{\enumerate}{\vspace{-4pt}\oldenumerate\setlength{\itemsep}{1pt}\setlength{\parskip}{0pt}\setlength{\parsep}{0pt}}

\newcommand*{\eps}{\varepsilon}

\def\beq{\begin{equation}}
\def\eeq{\end{equation}}
\def\bsp#1\esp{\begin{split}#1\end{split}}
\newcommand{\be}{\begin{equation}}
\newcommand{\ee}{\end{equation}}
\newcommand{\bea}{\begin{eqnarray}}
\newcommand{\eea}{\end{eqnarray}}

\def\spa#1.#2{\left\langle#1\,#2\right\rangle}
\def\spb#1.#2{\left[#1\,#2\right]}

\def\nn{\nonumber}

\newcommand{\F}{\mathfrak{F}}

\def\eps{\epsilon}

\def\ksl{\not{\hbox{\kern-2.3pt $k$}}}

\def\eps{\epsilon}

\def\bom#1{{\mbox{\boldmath $#1$}}}

\def\spa#1.#2{\left\langle#1\,#2\right\rangle}
\def\spb#1.#2{\left[#1\,#2\right]}
\def\lor#1.#2{\left(#1\,#2\right)}
\def\sand#1.#2.#3{%
\left\langle\smash{#1}{\vphantom1}^{-}\right|{#2}%
\left|\smash{#3}{\vphantom1}^{-}\right\rangle}

\def\F{ {\bom f}_{\! P}}

\definecolor{airforceblue}{rgb}{0.36, 0.54, 0.66}
\definecolor{bananayellow}{rgb}{1.0, 0.88, 0.21}
\definecolor{bittersweet}{rgb}{1.0, 0.44, 0.37}
\definecolor{blue(ncs)}{rgb}{0.0, 0.53, 0.74}
\definecolor{bole}{rgb}{0.47, 0.27, 0.23}
\definecolor{brass}{rgb}{0.71, 0.65, 0.26}
\definecolor{bronze}{rgb}{0.8, 0.5, 0.2}
\definecolor{brgreen}{rgb}{0.0, 0.26, 0.15}
\definecolor{burgundy}{rgb}{0.5, 0.0, 0.13}
\definecolor{cherry}{rgb}{1.0, 0.72, 0.77}
\definecolor{cocao}{rgb}{0.82, 0.41, 0.12}
\definecolor{citrine}{rgb}{0.99, 0.82, 0.07}

\makeatletter\DeclareRobustCommand*{\bfseries}{\not@math@alphabet\bfseries\mathbf\fontseries\bfdefault\selectfont\boldmath}\makeatother

\definecolor{darkred}{rgb}{0.5,0.0,0.0}

\definecolor{ultramarine}{rgb}{0.0,0.28,0.68}

\title{\boldmath 
Four-dimensional differential equations for the leading divergences of dimensionally-regulated loop integrals} 
\preprint{MPP-2022-137,USTC-ICTS/PCFT-22-31}

\author[a]{Johannes Henn} % \note{Also at Some University.}}
\author[b,c]{Rourou Ma} % \note{Also at Some University.}}
\author[d,e]{Kai Yan} % \note{Also at Some University.}}
\author[b,c]{Yang Zhang}

\affiliation[a]{Max-Planck-Institut f\"ur Physik, Werner-Heisenberg-Institut, D-80805 M\"unchen, Germany}
\affiliation[b]{Interdisciplinary Center for Theoretical Study, University of Science and Technology of China,
Hefei, Anhui 230026, China}
\affiliation[c]{Peng Huanwu Center for Fundamental Theory, Hefei, Anhui 230026, China}
\affiliation[d]{INPAC, Shanghai Key Laboratory for Particle Physics and Cosmology, School of Physics and Astronomy, Shanghai Jiao Tong University, Shanghai 200240, China}
\affiliation[e]{Key Laboratory for Particle Astrophysics and Cosmology (MOE), Shanghai 200240, China}

\emailAdd{henn@mpp.mpg.de}
\emailAdd{marr21@mail.ustc.edu.cn}
\emailAdd{yan.kai@sjtu.edu.cn}
\emailAdd{yzhphy@ustc.edu.cn}

\abstract{We invent an automated method for computing the divergent part of Feynman integrals in dimensional regularization.
Our method exploits simplifications from four-dimensional integration-by-parts identities. Leveraging algorithms from the literature,
we show how to find simple differential equations for the divergent part of Feynman integrals that are free of subdivergences.
We illustrate the method by an application to heavy quark effective theory at three loops.
}

\begin{document} 
%\graphicspath{{./Figures/}} %For Inkscape graphics
\maketitle
%\flushbottom

\section{Introduction}

Our ability to evaluate Feynman integrals is crucial for computing perturbative results in quantum field theory. This subject has a long and rich history of various methods being developed \cite{Smirnov:2012gma}.
In recent years, the method of differential equations \cite{Kotikov:1990kg,Bern:1992em,Remiddi:1997ny,Gehrmann:1999as} 
has become the main method for evaluating loop integrals that are needed for collider phenomenology. 
In particular, insights about writing the differential equations in a canonical form \cite{Henn:2013pwa}, has allowed both for an unprecedented degree of automation of the calculations, and has significantly pushed the boundaries of what is achievable. 

Calculations are typically done within dimensional regularization, with $D=4-2\eps$, and the results are expanded in a Laurent series in the dimensional regulator $\eps$. One advantage of the differential equations method is at the same time a drawback: in principle, the differential equations describe the functions under consideration at {\it any} order in $\eps$. In practice, however, one often wishes to know results up the the finite part only, so that a lot of unnecessary information is kept at intermediate steps. What is more, the higher-order terms in $\eps$ are typically scheme-dependent, which means that they are more complicated than the physically relevant terms.
Therefore it is desirable to develop methods that directly compute the physically relevant information, and profit from the simplifications.

Such a method exists for the case of (infrared- and ultraviolet-)finite Feynman integrals \cite{Caron-Huot:2014lda}. The authors showed that it is possible to directly compute integration-by-parts-relations (IBP) in four dimensions, taking however special care about possible contact terms that relate integrals of different loop orders. There are several advantages to this. Firstly, the restriction of finite integrals means that significantly fewer integrals and IBP relations between them need to be considered, and the resulting number of master integrals is also less than in the conventional $D$-dimensional case.  The simplifications in terms of handling the IBP relations in \cite{Caron-Huot:2014lda} are so substantial  that even certain three-loop four-point integrals with massive internal lines could be computed that were beyond the reach of conventional techniques. Secondly, when truncating to four dimensions, the canonical differential equations matrix takes a simple, `block-diagonal' structure, which reflects transcendental weight of the different Feynman integrals. A similar decoupling of the differential equations was also noticed in \cite{Tancredi:2015pta}. This makes it possible to iteratively solve for the master integrals, with increasing transcendental weight.

In many practical situations however, one often has to deal with divergent Feynman integrals.
In the context of dimensional regularization, one is really interested in the coefficient functions appearing in the Laurent series in the dimensional regulator $\eps$. 
In practice, only a finite number of terms is needed. Therefore it would be desirable to have a method that directly computes those finite coefficient functions.
Ideally, in order to profit from the advances mentioned above, this new method should be compatible with the canonical differential equations approach.
Our aim is to develop such a method.

As a first step in this direction, in this paper we focus on Feynman integrals that are free from subdivergences, i.e. that in dimensional regularization only diverge as $1/\eps$, and we develop a four-dimensional integration-by-parts and canonical differential equation method for the coefficient function. 
We leave the generalization to the general case for future work.
One might initially think that most Feynman diagrams have subdivergences, which limits the scope of the present paper. However, the structure of both ultraviolet (UV) and infrared divergences (IR) is very well understood, e.g. in terms of the famous Hopf algebra of renormalization \cite{Connes:1998qv}, and in terms of factorization properties (see \cite{Catani:1996vz,Dixon:2008gr,Almelid:2015jia} and references therein).
Therefore, at least in principle, one could use our method on Feynman integrals in those cases as well, for some suitably-defined subtracted integrals (see e.g. \cite{Anastasiou:2018rib} for recent work on an infrared subtraction for Feynman integrals). 

%Examples can be infrared-divergence scattering amplitudes, or ultraviolet divergences from renormalization. 
%Although there are various tricks for handling these divergences, and for formulating them in different ways, as far as we know there is no general purpose method that allows to compute them using simplified integration-by-parts relations. This motivates the present study.

We find it useful to focus on a particular type of divergences to present the new method. 
In this paper we focus on integrals in heavy quark effective theory (HQET). 
These integrals arise for example when massive particles emit soft radiation and capture infrared divergences associated to this radiation.
Equivalently, they can be used to compute ultraviolet divergences of certain Wilson lines with cusp (related to the angle-dependent cusp anomalous dimension).

Working with these integrals as a case in point, we explain the main ingredients of our method.
We first define HQET integrals that are free of subdivergences, and specify their leading divergence. 
We call the latter $I(\phi)$. Our goal is to compute $I(\phi)$.
The main advantage of focusing on this coefficient, as opposed to the full Laurent expansion is
that we can disregard unnecessary information.
We explain how one obtains four-dimensional integration-by-parts (IBP) identities valid for $I(\phi)$.
The master integrals appearing in these relations are all free of subdivergences of themselves,
and there are less of them compared to the generic case.  In this work,
we develop a new syzygy IBP method to forbid integrals with soft
divergence. (The original syzygy IBP method was developed to reduce
an IBP system's size
\cite{Gluza:2010ws, Schabinger:2011dz}.)

%\jmh{Here we can mention some words on how this is achieved (syzygy method), and mention the relationship to 
%what is in the literature (Kosower, Gluza).}

Moreover, we find, as in \cite{Caron-Huot:2014lda}, that there are integral identities that related 
different loop orders. As a result, the lower-loop information can be recycled.
Using these simplified IBP relations, we obtain differential equations for the master integrals.
Finally, we leverage ideas \cite{Dlapa:2020cwj} of how to transform these equations into a simple canonical form.

%\jmh{Some work still needed here.}

This paper is organized as follows: in the section \ref{sec:HQET}, we briefly review the Feynman integrals from web diagrams in HQET. In the section \ref{sec:integral_relation}, our method of generating IBPs for the divergent part of Feynman integrals, based on graded IBP operators and syzygy, is introduced. In the section \ref{sec:4D_initial}, we invent a four-dimensional version of the {\sc initial} algorithm \cite{Dlapa:2020cwj}, to generate canonical differential equations without $\epsilon$. %Then %several examples of our method, on the computation of two-loop and three-loop HQET integrals' divergent part, 
In section \ref{sec:3loop}, we provide a three-loop application of our method. Then we provide the summary and the outlook of our new method in the section \ref{sec:summary}. 
%The technical detail of systematically generating relations for integrals' divergent parts would be demonstrated by the appendix \ref{sec:techical_example_IBP}. 
% JMH: What is the additional information we give in the Appendix, as opposed to the main section? If something important is missing in the main section, can we explain it?

\section{From dimensionally-regularized Feynman diagrams to finite functions}
%\section{Four-dimensional Feynman integrals from web diagrams}
\label{sec:HQET}

Let us define the class of Feynman integrals that we study in this paper. 
%To study this problem 
We focus on integrals relevant for describing soft divergences occurring in the scattering of massive particles. 
These divergences can be described by an eikonal approximation, which leads to Wilson line correlators in position space.
Equivalently, in momentum space, one obtains heavy-quark-effective theory (HQET) integrals \cite{Grozin:2015kna}.

\subsection{Cusped Wilson lines in position space}
\label{subsec:cuspedWL}

\begin{figure}[h] 
    \centering 
    \begin{minipage}{0.4\linewidth}
        \centering
        \includegraphics[width=5cm,height=5cm]{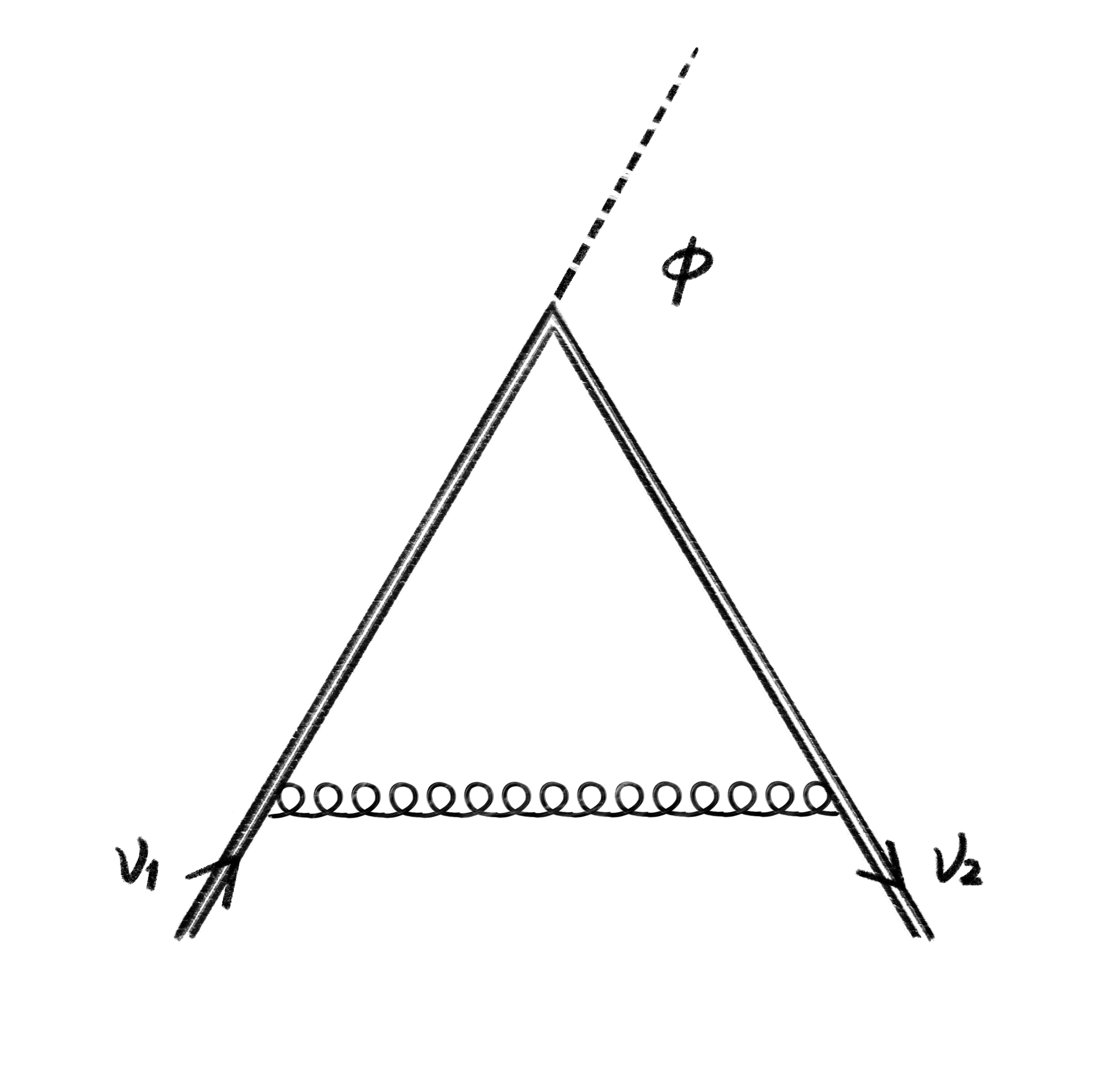} 
        \caption*{(a)}
    \end{minipage}
    \begin{minipage}{0.4\linewidth}
        \centering
        \includegraphics[width=5cm,height=5cm]{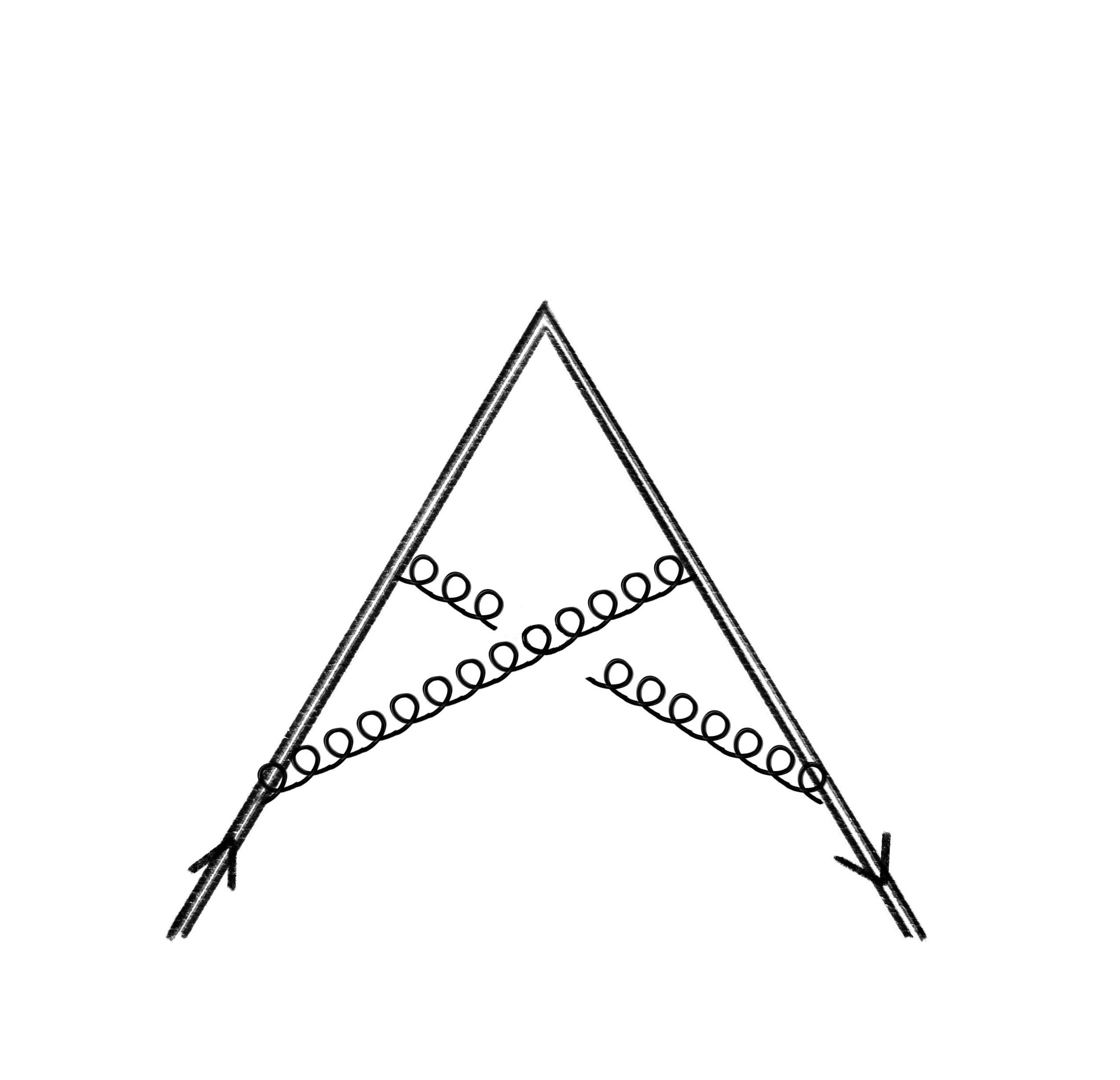}  
        \caption*{(b)}
    \end{minipage}
\caption{(a): One-loop Wilson line / heavy-quark effective theory diagram. (b) Example of a two-loop web diagram which has an overall divergence only.}
\label{fig:intro-cusp1}
\end{figure}
Consider a soft exchange in the scattering of two massive particles. 
The leading divergence for such a process is given by the eikonal diagram in Fig.~\ref{fig:intro-cusp1}(a). 
It depends on the scattering angle $\phi$, and is given by
\begin{align}
I^{(1)}(\phi, \eps)  \sim \frac{1}{\eps}   \phi \cot \phi + {\cal{O}}(\eps^0) \,,\qquad {\rm where} \quad \cos \phi = \frac{v_1 \cdot v_2}{\sqrt{v_1^2 v_2^2}} \,.
\end{align}
In the following we set $v_1^2 = v_2^2 = 1$ without loss of generality.
The coefficient of the $1/\eps$ pole is the angle-dependent cusp anomalous dimension \cite{Polyakov:1980ca,Brandt:1981kf,Korchemsky:1987wg}.
Our goal is to compute the coefficient for this and similar higher-loop diagrams directly, using four-dimensional methods, without having to deal with the higher-order terms in $\eps$.

When extracting information from divergent Feynman integrals it is paramount to consider carefully whether the computational steps are consistent and well-defined. 
In principle, there are several sources of divergences in Feynman integrals, in particular ultraviolet (e.g. from renormalization of the Lagrangian or of composite operators) and infrared (soft and collinear divergences associated to massless particles). Here we will focus on Feynman integrals that have one type of divergence only, leaving the case of integrals with divergences in several simultaneous regions for future work.

The knowledge of what region of loop integration produces the divergences allows us to extract its coefficient in a well-defined way.
Let us now illustrate this for the case of the one-loop integral shown in Fig.~\ref{fig:intro-cusp1}, and then explain how we approach the problem in general. In position space, the integral of Fig.~\ref{fig:intro-cusp1}(a) is given by
\begin{align}
I^{(1)}(\phi, \eps)  \sim \int_0^\infty d\tau_1 d\tau_2 \frac{v_1\cdot v_2} {[(\tau_1 v_1+\tau_2 v_2)^2]^{1-\eps}} e^{-\tau_1 -\tau_2}\,.
\end{align}
Here $e^{-\tau_1 -\tau_2}$ is a cutoff that makes sure that the integral gives only the divergence at $\tau_1, \tau_2 \sim 0$ that is of interest to us.\footnote{In fact, we are extracting the UV divergence of this integral. This may seem strange as our starting point was a soft exchange, i.e. an IR effect. The explanation is that these UV and IR divergence are related $\Gamma_{\rm UV} = - \Gamma_{\rm IR}$, via the formal relation $\int_0^\infty d\tau/\tau^{1-\eps} =0$.} 
We can make the latter manifest by an overall rescaling,
\begin{align}\label{eq:intro-example1}
I^{(1)}(\phi, \eps)  \sim \underbrace{\int_{0}^{\infty}  \frac{d\rho}{\rho^{1-\eps}}  
e^{-\rho}}_{1/\eps +{\cal O}(\eps^0)} 
\times  \underbrace{\int_0^1 dz \frac{v_1\cdot v_2} {[(z v_1+ (1-z) v_2)^2]^{1-\eps}}}_{\phi \cot \phi + {\cal O}(\eps^0)} \,.
\end{align}

\subsection{HQET integrals in momentum space}
\label{subsec:HQET}

One can equivalently write the integral discussed in subsection \ref{subsec:cuspedWL} in HQET language in momentum space.
We briefly recall some notations of HQET (see \cite{Grozin:2004yc} for a detailed review). 
HQET is an effective theory describing the dynamics of heavy quarks.
In the heavy mass limit, the heavy quark's propagator becomes
\begin{equation}
    i\frac{\slashed p+m}{p^2-m^2 + i \eta} \to i \frac{1+\slashed v}{2 v\cdot k+i\eta}\,,
    \label{eq:HQET_propagator}
\end{equation}
where $p=mv+k$. $v$ is the classical velocity of the heavy quark, normalized as $v^2=1$, while $k$ is its off-shell momentum. In this limit, the mass parameter factorizes out and the quantum field theory computation is simplified significantly. 
In practice, we may use an infrared regulator $\delta$ for the linear propagator in \eqref{eq:HQET_propagator}. 
Since $\delta$ is the only scale for the HQET integrals, it is safe to set $\delta=1$.\footnote{Another subtlety is that frequently in the literature, a minus sign is added on the linear propagator and the denominator reads $-2v\cdot k +\delta-i\eta$.}
In HQET language, diagram Fig.~\ref{fig:intro-cusp1}(a)  is given by (up to an overall factor, $D=4-2\eps$)
\begin{align}\label{eq:intro-example2}
I^{(1)}(\phi, \eps) \sim \int \frac{d^{D}k}{i \pi^{D/2}} \frac{1}{k^2 (2 k\cdot v_1 + 1)(2 k\cdot v_2 +1)} \,.
\end{align}
In eq. (\ref{eq:intro-example2}), the divergences comes from an overall rescaling where $k^\mu \sim \infty$.
Now, in the above example, one could directly compute the coefficient of the $1/\eps$ pole by setting $\eps=0$ in the second factor of eq. (\ref{eq:intro-example1}).
Something similar can be done in momentum space.
We prefer however to develop a formalism that can be applied directly for covariant, four-dimensional Feynman integrals. 
The reason for this is that in this way we obtain a more general setup that we expect may be used for larger classes of Feynman integrals.
For this reason we define
\begin{align}\label{eq:intro-example3}
J^{(1)}(\phi) = \lim_{\eps \to 0} \left[ \eps\, I^{(1)}(\phi, \eps) \right] \,.
\end{align}
Our goal is then to derive four-dimensional IBP relations that are valid for $J$. 

\subsection{Higher-loop diagrams free of subdivergences}
\label{subsec:webs}

Let us discuss the generalization to higher loops. Again, we focus on Feynman integrals with the divergence coming from one region of loop integration. In the case of our HQET integrals, we assume that they are free of subdivergences, and that the overall divergence is associated to a region where all loop momenta become large. See Fig.~\ref{fig:intro-cusp1}(b) for a two-loop example.
Fig.~\ref{fig:intro-cusp1}(b) is an example of a so-called web diagram \cite{Frenkel:1984pz,Gatheral:1983cz}.
% are these the correct references for webs?
 By definition, webs do not have one-particle irreducible subdiagrams, which ensures the absence of subdivergences. Web diagrams occur naturally in the study of Wilson lines in the context of non-Abelian eikonal exponentiation \cite{Frenkel:1984pz,Gatheral:1983cz}. In a nutshell, they provide a setup for computing the cusp anomalous dimension from Feynman diagrams without subdivergences. 
% Strictly speaking, we could have subdivergences from the renormalization of the Lagrangian, for example. Shall we explain that we do not consider them.
So they are ideal objects to study for our purposes. To extract the overall divergence, we define in analogy to eq. (\ref{eq:intro-example3}),\footnote{%
It is also possible define $\tilde J^{(L)}(\phi) =\lim_{\eps \to 0} \left[  \eps L \, I^{(L)}(\phi, \eps) \right]$. The factor of $L$ in this definition  occurs naturally from dimensional regularization, and will facilitate comparing integrals at different loop levels. See subsection \ref{subsection:cross-loop} for the details.
}
%\begin{align}\label{eq:intro-example4}
%J^{(L)}(\phi) = \lim_{\eps \to 0} \left[ L \eps \, I^{(L)}(\phi, \eps) \right] \,.
%\end{align}
\begin{align}\label{eq:intro-example4}
J^{(L)}(\phi) = \lim_{\eps \to 0} \left[  \eps \, I^{(L)}(\phi, \eps) \right] \,.
\end{align}

In this paper we focus on HQET web diagrams built from two eikonal lines (corresponding to the heavy quarks), and an arbitrary number of massless exchanges. 
 We impose the following conditions on them:
 %These integral are the $L$-loop Feynman integrals for the eikonal diagrams (triangular diagrams with two heavy quark lines) in HQET with the property that:
\begin{enumerate}
  \item \label{condition1} The integrand has the overall scaling dimension $-4L$, for the transformation $k_i \to \lambda k_i$, $i=1,\ldots, L$. This follows from eikonal Feynman rules. 
 % In other words, the integrals should have the global UV logarithmic divergence.
  \item \label{condition2}  No UV divergence in any subloop subdiagram. % of these Feynman integrals.
\item \label{condition3}  No infrared (IR) divergences. This forbids higher powers of massless propagators. %There is no double gluon propagator in these Feynman integrals, due to the absence of soft divergence.
\end{enumerate}
We call such integrals ``admissible".
It follows from these properties that in dimensional regularization, admissible integrals have an overall $\epsilon^{-1}$ divergence only.
We are interested in the coefficient of the divergence in (\ref{eq:intro-example4}).
%{\color{red} By the abuse of notation, we call integrals  satisfying the conditions above ``admissible" integrals in this paper.} In the dimensional regularization scheme, these integrals has the $\epsilon^{-1}$ divergence. As mentioned in the introduction, we are interested in the $O(\epsilon^{-1})$ part of the admissible integrals,
%\begin{equation}\label{defleadingdivergence}
%    J^{(L)} = \lim_{\eps \to 0} \left[  \eps \, I^{(L)} \right] \,.
%\end{equation}
%where $I^{(L)}$ is an admissible integral. 
%The kinematic variable is,
%\begin{equation}
 %   v_1 \cdot v_2 =\cos \phi=\frac{1}{2}(x+\frac{1}{x})\,
%\end{equation}
%where $v_i$, $i=1,2$ are the velocities of the two heavy quark lines.  

\subsection{Finding admissible integrals}
\label{subsec:admissble}

\begin{figure}[H] 
    \centering 
    \begin{minipage}{0.32\linewidth}
        \centering
        \includegraphics[width=4.5cm,height=4.5cm]{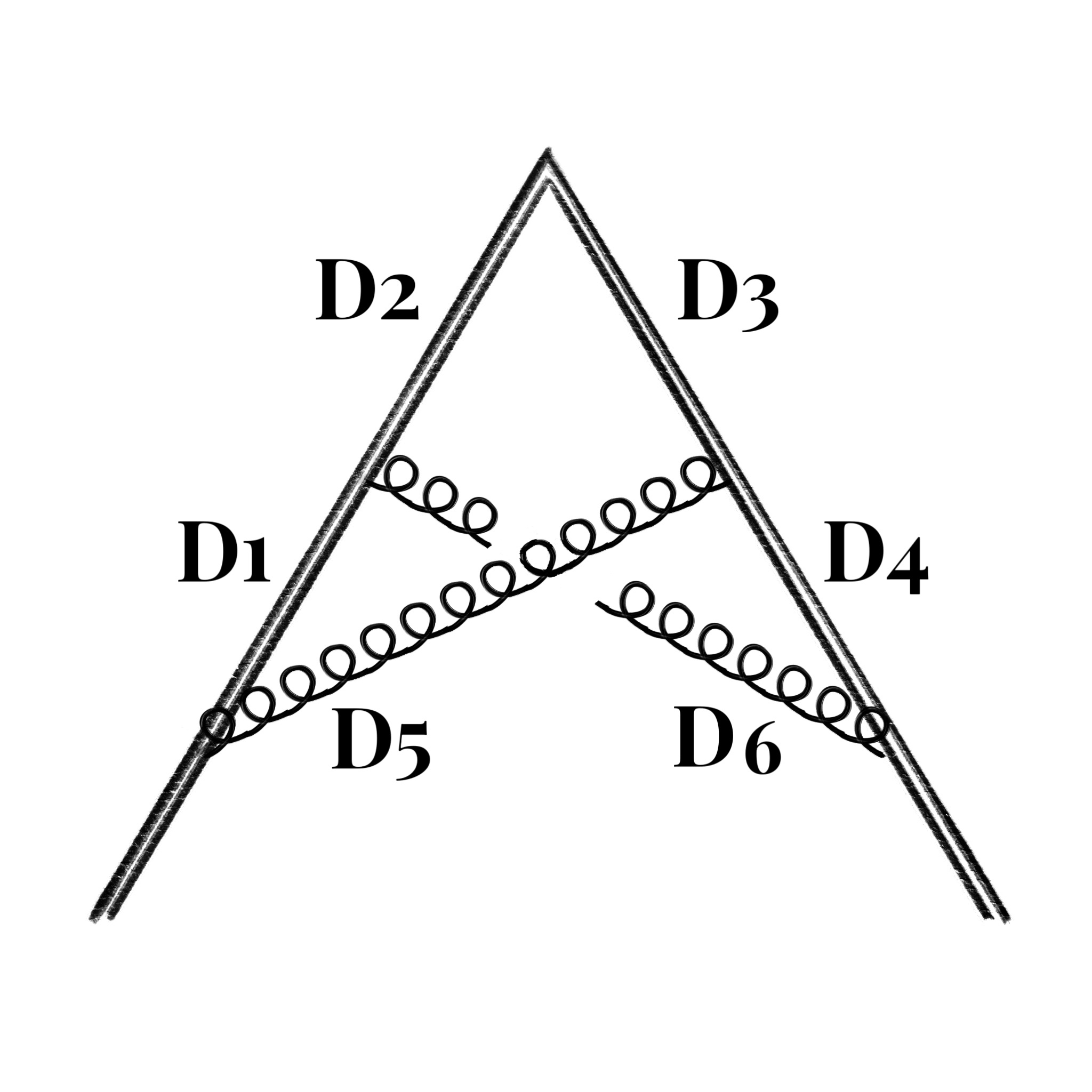} 
        \caption*{$A_1$}
    \end{minipage}
    \begin{minipage}{0.32\linewidth}
        \centering
        \includegraphics[width=4.5cm,height=4.5cm]{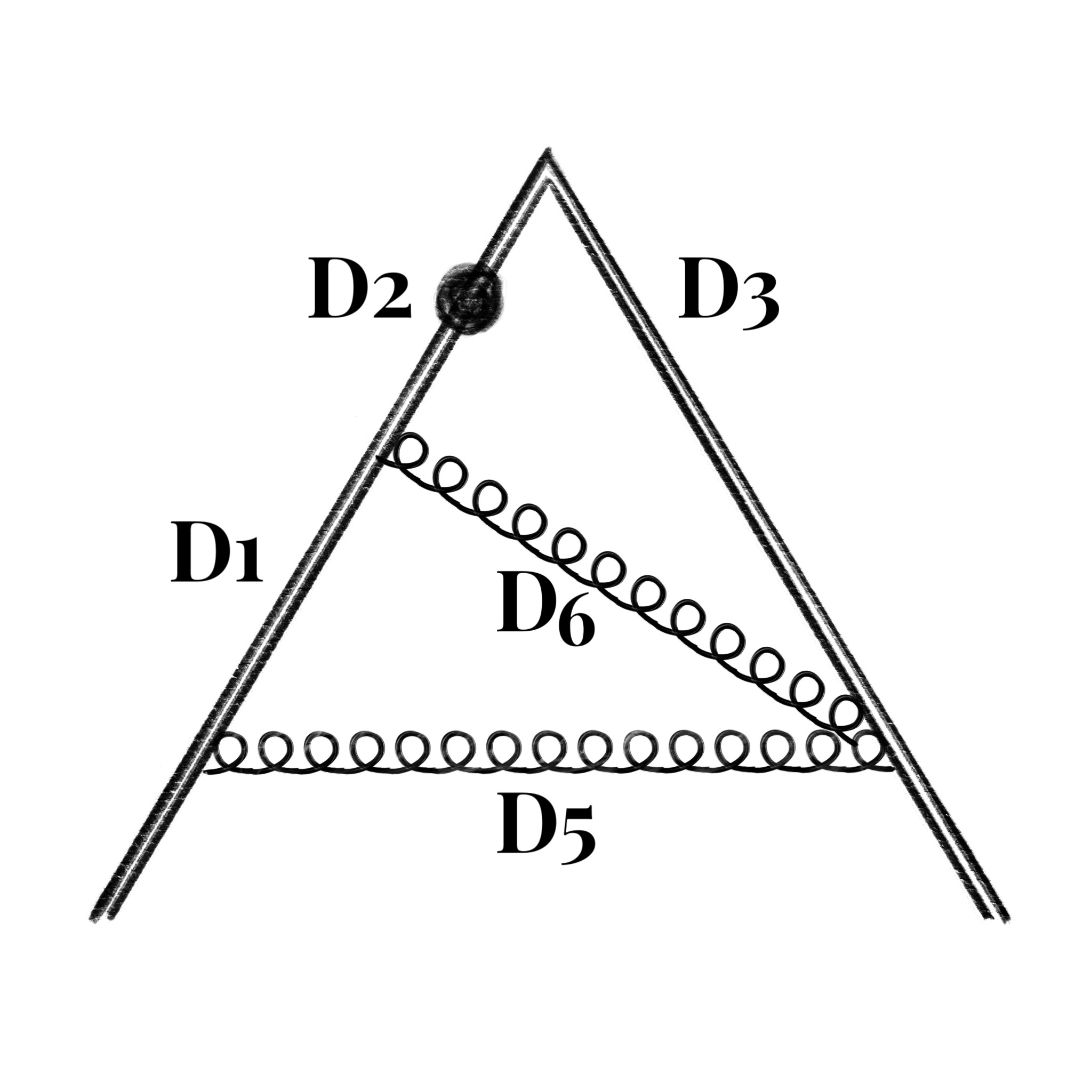}  
        \caption*{$B_1$}
    \end{minipage}
    \begin{minipage}{0.32\linewidth}
        \centering
        \includegraphics[width=4.5cm,height=4.5cm]{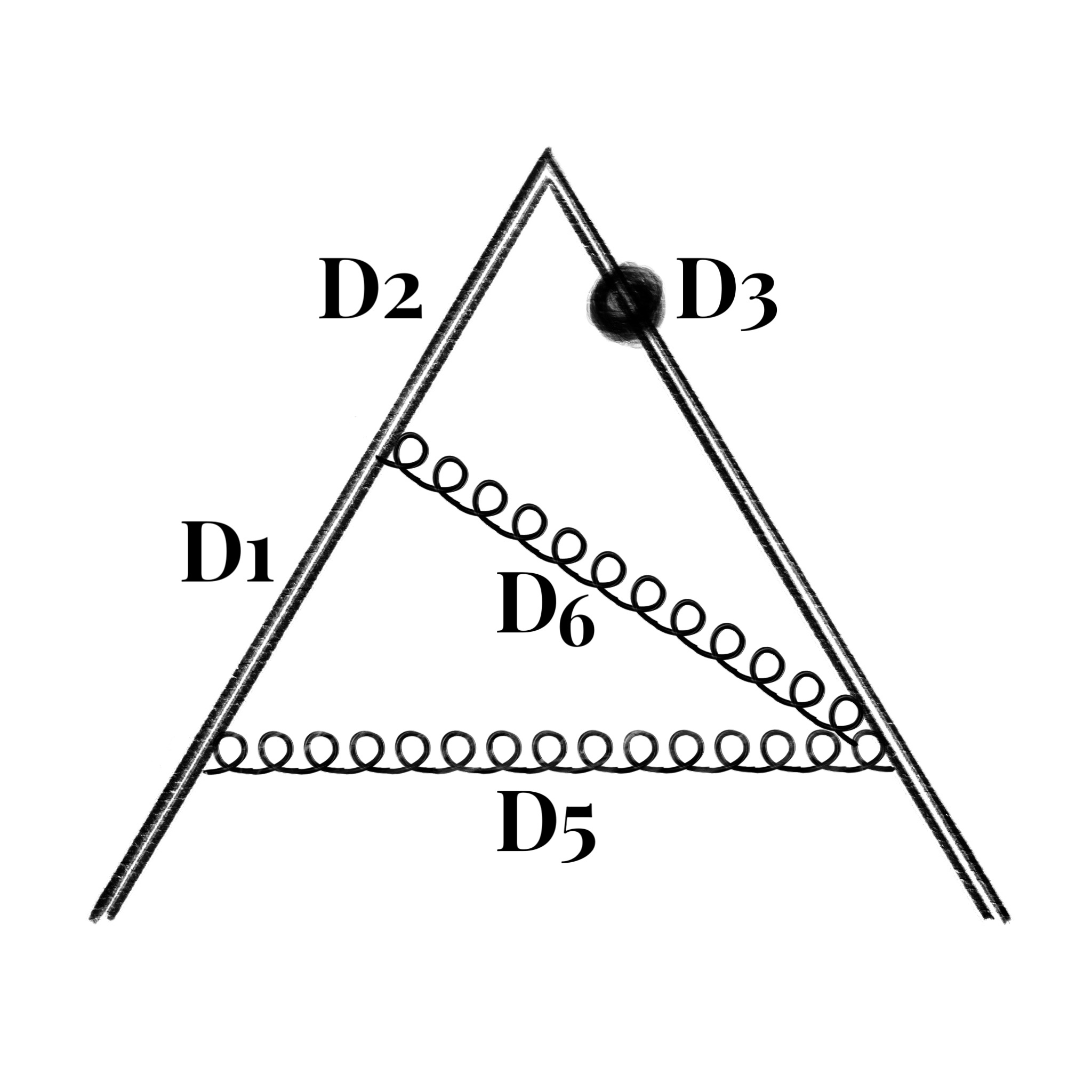}  
        \caption*{$B_2$}
    \end{minipage}

	\begin{minipage}{0.32\linewidth}
        \centering
        \includegraphics[width=4.5cm,height=4.5cm]{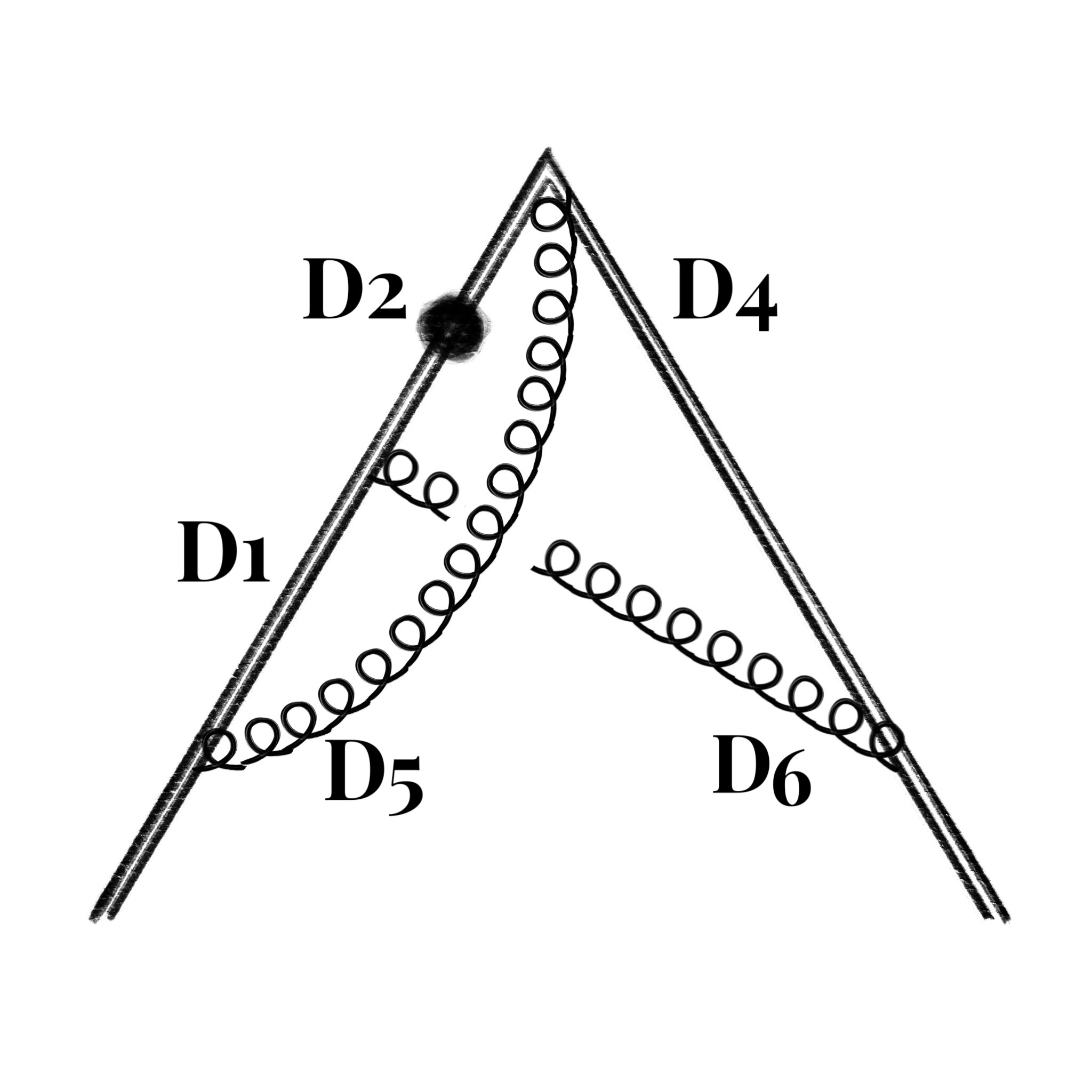} 
        \caption*{$B_5$}
    \end{minipage}
    \begin{minipage}{0.32\linewidth}
        \centering
        \includegraphics[width=4.5cm,height=4.5cm]{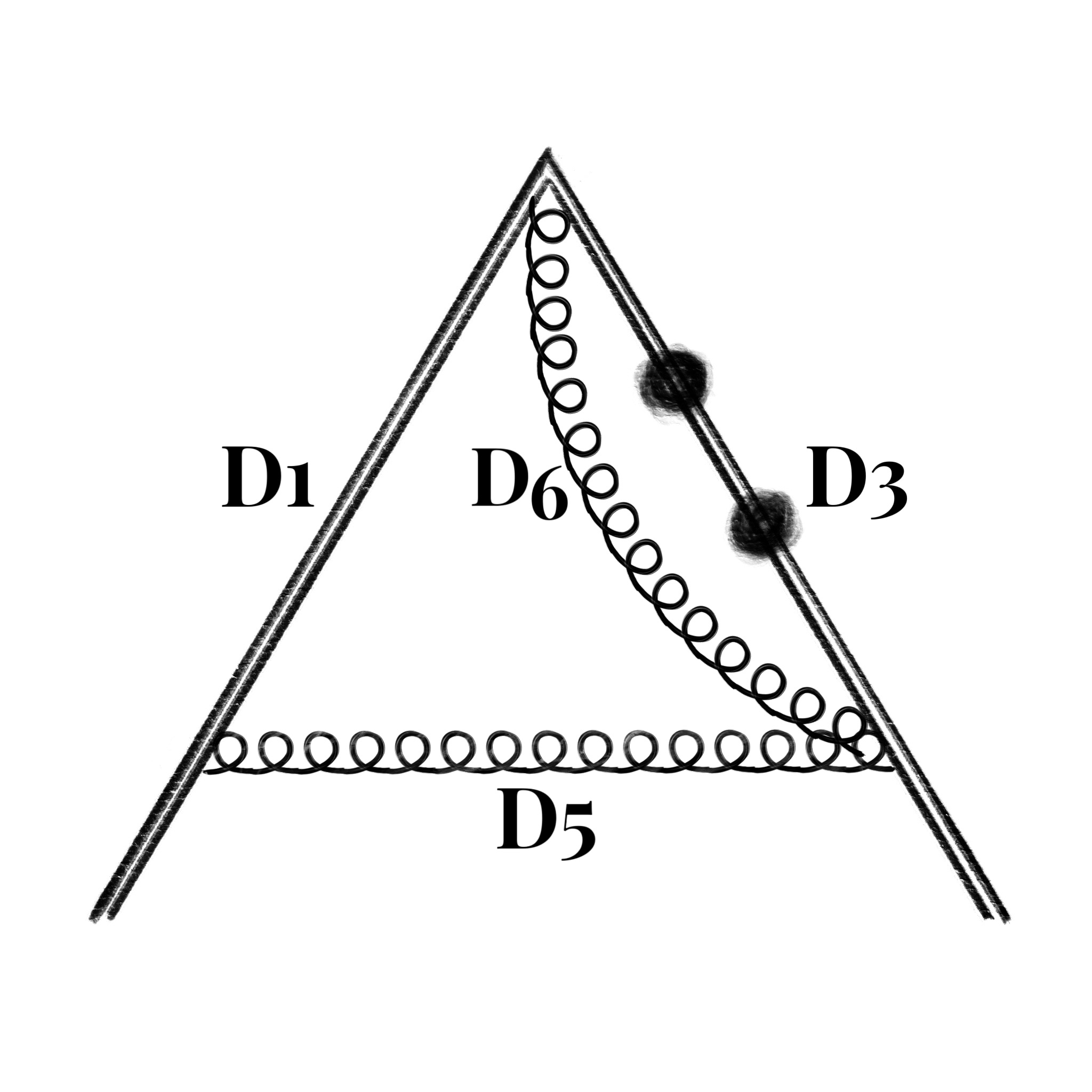}  
        \caption*{$C_1$}
    \end{minipage}
    \begin{minipage}{0.32\linewidth}
        \centering
        \includegraphics[width=4.5cm,height=4.5cm]{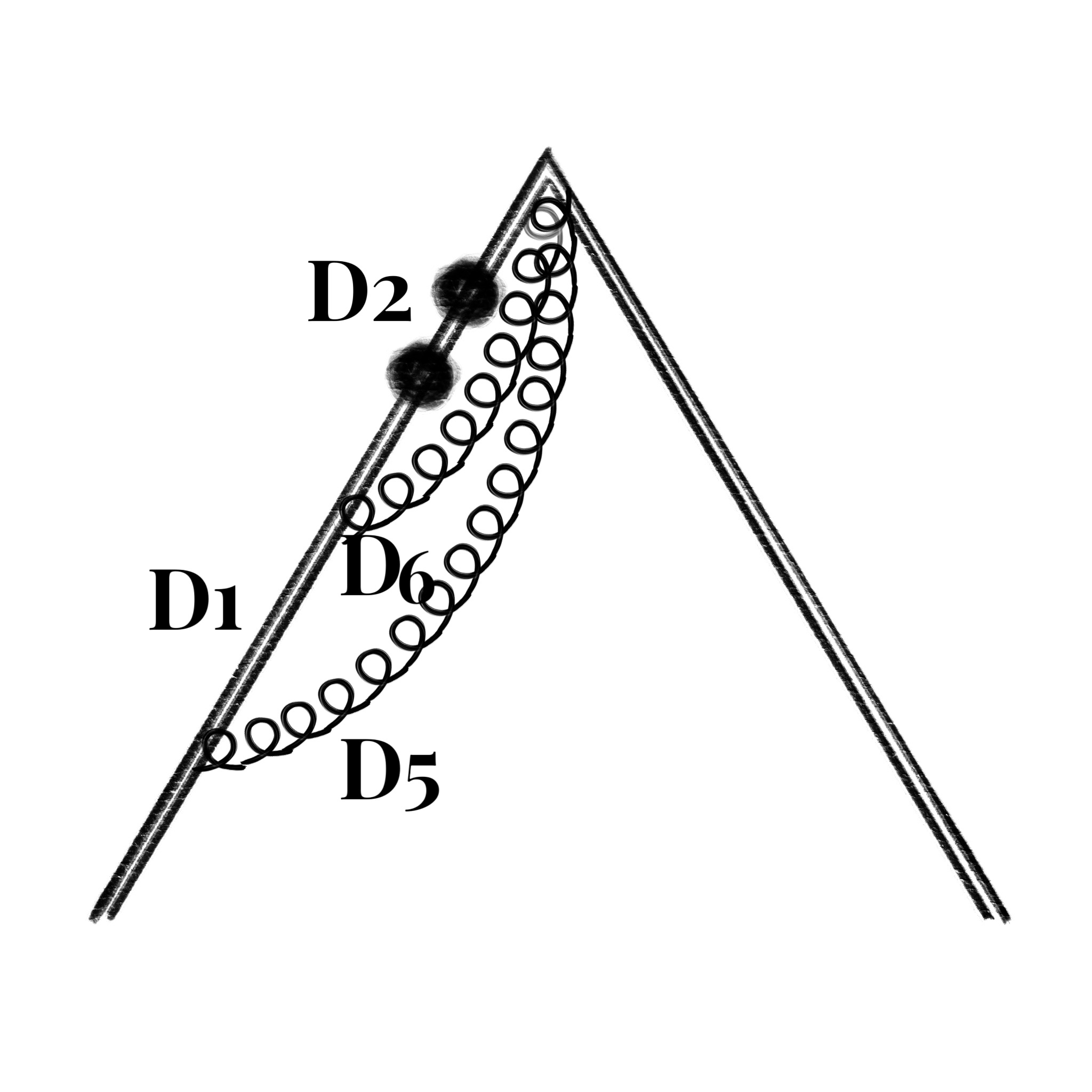}  
        \caption*{$C_4$}
    \end{minipage}
%\end{figure}
%
%\begin{figure}[H]    
%    \begin{minipage}{0.64\linewidth}
   %     \centering
      %  \includegraphics[width=8.4cm,height=4.2cm]{FeynmanDiagram/C1}  
      %  \caption*{$C_1$}
%    \end{minipage}
%    \begin{minipage}{0.32\linewidth}
%        \centering
%        \includegraphics[width=4.2cm,height=4.2cm]{FeynmanDiagram/D1}  
%        \caption*{$D_1$}
  %  \end{minipage}
\caption{\label{fig:admissiblecrossedladderintegrals}  Admissible diagrams in the crossed ladder integral family. The dot stands for a doubled propagator. Further admissible integrals that involve numerator factors are given in the main text.   %\ky{I have replaced the notation $D_i$ for admissible integrals with $C_i$, to avoid confusion with the notation for propagators.} 
%\jmh{Figure for $B_5$ is missing}.  %\jmh{Remove identities to one-loop integrals (just show the two-loop integrals $C_1, D_1$.
}
%\ky{$D_1$ is not an admissible integral.}
%\jmh{Add two (?) other diagrams. It would be nice to number the propagators in the crossed ladder figure.}
% }
\end{figure}

The three conditions 
%\ref{condition1},\ref{condition2},\ref{condition3} 
formulated in subsection \ref{subsec:webs} can be translated to constraints on the Feynman integrals' indices. 
While this restricts the space of integrals of interest to us, in general the solution space is infinite dimensional. 
In practice it is sufficient to work with a finite-dimensional subspace. We do this by placing further constraints on degrees of numerators or denominators. 
Then we can find all admissible integrals are subject to these additional conditions with the help of computer algebra.

%So in principle, we can find all admissible integrals, with certain bounds for the numerator and denominator degrees, 
%of a given diagram computer algebra. 

%However, the second condition requires some explanations since it has a subtlety with the usual notation of irreducible scalar products (ISPs).  
%Usually, given an $L$-loop Feynman integrals with $E$ external legs and $n$ propagators, one adds $LE+\frac{1}{2} L(L+1)-n$ ISPs for the computation. The propagators and ISPs are linearly independent. Any other scalar products would be dependent of the propagators and ISPs, and hence called reducible scalar products (RSPs). RSPs can be removed in general computations \cite{Ossola:2006us,Mastrolia:2006ki,Ellis:2008ir,Ellis:2011cr,Mastrolia:2011pr,Zhang:2012ce,Mastrolia:2012an,Mastrolia:2012wf}. However, for our purpose, we find RSPs  useful, because some admissible integrals can be neatly written with the help of RSPs, but have complicated expressions if  RSPs are reduced. Therefore, we consider RSPs for our searching of admissible integrals. (See ref.~\cite{Chen:2022jux} for the recent application of RSPs). This subtlety is demonstrated in the following example.
%\jmh{This is purely a notational convention, I would not call it a subtety.}

Let us illustrate this for the diagram shown in Fig.~\ref{fig:intro-cusp1}(b).
We define the crossed-ladder integral family ${\rm T^{(2)}}[a_1, \cdots a_7]$ as follows.
\begin{align} \label{CLdef}
 & \qquad{ \rm T^{(2)}}[a_1,\cdots,a_7] =   \int \frac{d^{D} k_1 }{i \pi^{D/2}}  \frac{d^{D} k_2 }{i \pi^{D/2}}\,  \frac{ D_7^{-a_7}}{ D_1^{a_1} \,  \cdots\,  D_6^{a_6} }\,, \quad 
\end{align}
where
\begin{align}
D_1 &= -2 k_1 \cdot v_1 + \delta , \quad  D_2 = -2 (k_1 +k_2) \cdot v_1 + \delta, \quad  D_3 = -2 (k_1 +k_2) \cdot v_2 + \delta, \nn \\
 D_4 &= -2 k_2 \cdot v_2 + \delta, \quad   D_5 = -k_1^2, \quad  D_6 = - k_2^2,  \quad  D_7= k_1 \cdot k_2  \,. 
\end{align}
%In principle there could be an infinite number of admissible integrals. 

We formulate the following additional criterion (based on experience) for these two-loop integrals: if the number of propagators is $p$, then we allow up to $7-p$ doubled propagators.
For example, for six propagators, we allow one doubled propagator; for five propagators, we 
allow two doubled propagators, 
and so on. Of course, we could later relax this condition, if necessary. However we will see below that they are sufficient for our purposes.
We then find the following admissible integrals (are subject to the additional conditions).

In top sector with six propagators (i.e. indices $a_1, \ldots a_6$ positive), we have the scalar crossed ladder diagram (see Fig.~\ref{fig:admissiblecrossedladderintegrals})
\begin{align}
A_1 &= {\rm T^{(2)} } [1,1,1,1,1,1,0]\,. \label{CLAFin} 
\end{align}
Moreover, there are two additional admissible integrals that involve numerators. The first one is given by the scalar crossed ladder integrand, but with additional factor 
\begin{align}
\frac{-2 k_1 \cdot v_2}{D_1}  = \frac{D_3 - D_4}{D_1}\label{eq:decompreducible}
\end{align}
inserted. 
Because of eq. (\ref{eq:decompreducible}), this integral (and a similarly constructed one) can be written as a difference of integrals in subsectors, as follows
\begin{align}
A_2 &={\rm T^{(2)} }[2, 1, 0, 1, 1, 1, 0] -{\rm T^{(2)} }[2, 1, 1, 0, 1, 1, 0], \label{CLAFin2}\\
A_3 &={\rm T^{(2)} }[1, 0, 1, 2, 1, 1, 0]-{\rm T^{(2)} }[0, 1, 1, 2, 1, 1, 0] \label{CLAFin3}
\end{align}
However, it is important to realize that only the linear combinations in eqs. (\ref{CLAFin2}) and (\ref{CLAFin3}) are admissible integrals, and not the individual terms.
For this reason we find it more appropriate to think of these integrals as belonging to the top sector with six propagators. In order to make this manifest, we introduce the auxiliary noation
\begin{align}
{\rm t^{(2)}}[a_1,\cdots,a_9] \equiv   \int \frac{d^{D} k_1 }{i \pi^{D/2}}  \frac{d^{D} k_2 }{i \pi^{D/2}}\,  \frac{ D_7^{-a_7} D_8^{-a_8} D_9^{-a_9}}{ D_1^{a_1} \,  \cdots\,  D_6^{a_6} }\,.
\label{eq:RSP_integral}
\end{align}
where  we define two reducible scalar products (RSPs) $D_8 = -2 k_1 \cdot v_2, \; D_9 = -2 k_2 \cdot v_1$. In this way, admissible integrals in \eqref{CLAFin} can be neatly written as,
\begin{align} 
A_2 &={\rm  t^{(2)} } [2,1,1,1,1,1,0,-1,0], \\
A_3 &= {\rm t^{(2)} } [1,1,1,2,1,1,0,0,-1]\,.
\end{align}
%We remark that the use of RSPs is not only for simplifying notations, but later during the IBP reduction, it would ensure that every term is  admissible in all intermediate steps. This feature validates the IBP reduction algorithm for the Feynman integral's divergent part.
A final comment is in order. There are further integrals related to the above by symmetry relations (e.g. the $v_1 \leftrightarrow v_2$ symmetry implies a flip symmetry of graphs), which we do not show.
One could generate these additional integrals, and then eliminate them automatically with the help of the integration-by-parts relations discussed in the next section.

For sectors with five propagators, we find
\begin{align}
B_1 &=   {\rm T^{(2)}}[1,2,1,0,1,1,0]   \\
 B_2 &=  {\rm T^{(2)}}[1,1,2,0,1,1,0]  \\  
 B_3 &= {\rm t^{(2)}}[2,1,2,0,1,1,0,-1,0] \,,\\
    B_4 &=  {\rm t^{(2)}}[2,2,1,0,1,1,0,-1,0] \,,\\
    B_5 &= {\rm T^{(2)}} [1,2,0,1,1,1,0]  \,. \\
%\end{align}
%{\ky{and}} 
%\begin{align} 
B_6 & = {\rm T^{(2)}[1,3,-1,1,1,1,0]} \\
B_7 & = {\rm t^{(2)}[1,2,0,2,1,1,0, 0, -1] } \\
B_8 & = {\rm t^{(2)}[2,2,0,1,1,1,0, -1, 0] } 
\end{align} 
For sectors with four propagators, we find
\begin{align} \label{T2triangle}
C_1 &= 
{\rm T^{(2)}} [1,0,3,0,1,1,0]\,,\\
%\end{align} 
%{\ky{and}} 
%\begin{align}
C_2 &=
{\rm T^{(2)}} [1,-1,4,0,1,1,0]\,,\\
C_3 &=
{\rm t^{(2)}} [2,0,3,0,1,1,0,-1,0]\,,\\
C_4 &= %2 \,
{\rm T^{(2)}} [1,3,0,0,1,1,0]\,.\\
C_5 &= 
{\rm T^{(2)}} [1,4,-1,0,1,1,0]\,.\\
C_6 &=
{\rm t^{(2)}} [2,3,0,0,1,1,0,-1,0] \,. 
\end{align} 
%\jmh{Is this list really complete? In text that I removed from the IBP section it seemed to me that more integrals were discussed (But I couldn't follow that discussion because it talked about sectors, not referring to integrals defined here and their pictures).} \ky{I am not sure if my old code to search for admissible  integrals has been tested or not. Can someone please double check if we have the complete list of admissible integrals with no more than $7-p$ dots ? } 
These integrals are shown in Fig.~\ref{fig:admissiblecrossedladderintegrals}. Since the integrals above are all admissible, the limit of eq. (\ref{eq:intro-example4}), that we are mostly interested in, is well defined.
From now on we will use the same notation both for the integral and its value in the limit of eq. (\ref{eq:intro-example4}), hoping that this does not lead to confusion.

\section{Four-dimensional integral reduction identities}
%\section{Four-dimensional integration-by-parts relations 
%from syzygies, and loop-order reduction relations}
\label{sec:integral_relation}

%Our goal is to develop a canonical differential equation method for the divergent part of integrals with only one region of divergence. We refer to the references \cite{Henn:2014qga,Lee:2014ioa,Meyer:2016slj,Meyer:2017joq,Gituliar:2017vzm,Prausa:2017ltv,Henn:2020omi,Chen:2020uyk,Dlapa:2021qsl,Dlapa:2022nct,Chen:2022lzr,Chen:2022fyw}, for developments of finding canonical differential equations for all $\epsilon$ orders of integrals.

%In order to set up a differential equation system, in this paper, 

Our next goal is to find linear relations between the $\epsilon^{-1}$ parts of
admissible integrals. We denote  ${\mathbb  V}$  as the linear
space spanned by admissible integrals. We then find a basis of this space, taking into account the following additional relations:
%After these preparations, we need to generate IBP relations for admissible integrals.
%We do this in the following way:
\begin{enumerate}
\item Apply the zero sector condition \cite{Lee:2012cn,Lee:2014ioa}
to $\mathbb V$ to remove vanishing integrals, and also the Pak algorithm \cite{Pak:2011xt} to identify equivalent sectors. This reduces the number of admissible integrals that need to be considered.
    \item Find IBP operators that generate relations between
      admissible integrals. Firstly, the operators need to be {\it
        graded}, which means that all terms have the same scaling
      dimension w.r.t. the loop momenta. This is necessary in order to
      be able to apply the UV conditions number \ref{condition1} 
      and
      \ref{condition2} in subsection \ref{subsec:webs}, for admissible integrals. Secondly, in order to
      fullfil the condition number
      \ref{condition3} in subsection \ref{subsec:webs}, the operators need to satisfy certain syzygy
      conditions like those in \cite{Gluza:2010ws,Schabinger:2011dz}. 
          \item For $L$-loop admissible integrals with bubble sub-diagrams, 
    use a reduction identity for integrating out the bubble,
    leading to an $(L-1)$-loop admissible integral. 
\end{enumerate}

\subsection{Graded IBP operators and the syzygy method}
\label{subsect:gradedIBP}

A generic IBP relation has the schematic form,
\begin{gather}
 \int  \bigg(\prod_{i=1}^L  \frac{d^{4-2\epsilon}
    k_i}{i\pi^{2-\epsilon}}\bigg) O\bigg(\frac{1}{D_1 ^{a_{1}} \ldots D_{n}^{a_{n}}} \bigg)=0
    \label{eq:IBP_generic}
\end{gather}
with a differential operator $O$,
\begin{equation}
  O=\sum_{i=1}^L \frac{\partial }{\partial k_i^\mu} u_i^\mu\,,
  \label{eq:IBP_operator}
\end{equation}
with vectors $u_i$. 
However, this generic form of IBP relations contains non-admissible integrals in general.
The reason is that the vectors $u_i$ may change the power counting of sub-loop diagrams and thus can introduce sub-loop UV divergences. 
Moreover, when $\partial /\partial k_i^\mu$ acts on gluon propagators, it increases the propagator power and may worsen the infrared properties. 
%In the other words, the left hand side of \eqref{eq:IBP_generic} may not consist of {\it admissible} integrals,
% or be in the linear space $ \tilde {\mathbb V}$. 
%and therefore the generic IBP relations are not directly applicable.
%
%It is possible to combine several IBP relations of the generic form, to get IBP relations for admissible integrals. 
%However, this approach deals with a huge number of IBP relations and thus is complicated. 
%Instead of using generic IBPs, we develop a new way of IBP relations with the following constraints:
%\jmh{Repetition of what was said at the beginning of the
%section. Maybe combine?}
Our goal is to generate valid IBP relations between admissible integrals. (Below we also use a similar strategy for finding differential equations between admissible integrals.)
In order to achieve this, we develop a new way of generating IBP relations with the following constraints:
\begin{enumerate}
    \item The IBP operator in \eqref{eq:IBP_operator} should be {\it graded}. This means that under the scaling of each loop momenta, $k_i \mapsto \lambda k_i$, $i=1,\ldots, L$, 
        \begin{equation}
            O \mapsto \lambda^{\beta_i} O,\quad \beta_i \in \mathbb Z
        \end{equation}
% Note that under this scaling 
%\begin{equation}
   %         \frac{\partial}{\partial k_i^\mu} \mapsto \lambda^{-1} \frac{\partial}{\partial k_i^\mu}  
%\end{equation}
The tuple $\beta_i$, $i=1,\ldots, L$ are the scaling dimensions for each individual loop momentum.  
An overall scaling dimension condition for IBP vectors has been used for the linear algebra algorithm of IBP relations without doubled propagators \cite{Schabinger:2011dz}. Here we use individual scaling dimensions to impose a well-defined power counting on the IBP differential operator for each sub loop. When deriving IBP relations we will use this power counting condition to ensure the absence of subdivergences. 
In practice,  it is easy to make an ansatz for graded IBP operators with given a scaling dimensions $\beta_i$, $i=1,\ldots, L$.

\item The IBP operator in \eqref{eq:IBP_operator} should not produce higher power of  gluon propagators. Let $D_k$ be a gluon propagator, then, the requirement reads,
\begin{equation}
    O D_k = g_k D_k
     \label{eq:syzygy_relation}
\end{equation}
where $g_k$ is a polynomial in the scalar products involving loop and
external momenta. This is the syzygy relation for IBP relations
\cite{Gluza:2010ws}. The original purpose of using syzygy relations
is to reduce the number of IBP relations for the reduction step. See
\cite{Schabinger:2011dz, Larsen:2015ped, Boehm:2017wjc, Boehm:2018fpv, Bendle:2019csk}
for the development of the syzygy method for IBP reduction. In this
paper, \eqref{eq:syzygy_relation} is imposed to control the soft
divergence and it is only on massless (gluon) propagators. Note that
although this condition is not imposed on every propagator, 
%but as a ``side effect'', 
it still significantly reduces the number of IBP relations.
\end{enumerate}
To find suitable IBP differential operators, we use these conditions together. It is natural to make an ansatz for graded differential operators (first condition) and then solving the syzygy relation  \eqref{eq:syzygy_relation} with the linear algebra algorithm in \cite{Schabinger:2011dz}. Given an ansatz for the graded IBP operator, \eqref{eq:syzygy_relation} becomes a sparse linear algebra problem. In practice, we wrote a proof-of-concept code in {\sc Mathematica} to find such differential operators, which we make available in the auxiliary file ``demo/graded\_vector\_demo.wl". There is also an available package {\sc NeatIBP} \cite{NeatIBP} to generate small-size IBP relations by syzygy method.

Then we apply such differential operators on Feynman integrals with suitable indices, to get relations between admissible integrals. It is notable that not every term in such a relation is admissible, but suitable combinations of terms are admissible. In practice, to find such combinations, we first set up an integrand basis of admissible integrals, and then expand the relation over this integrand basis. This step is powered by a finite-field linear algebra code, which depends on {\sc SpaSM} \cite{spasm}. We provide this code in the auxiliary file ``demo/3lHQET\_graded\_IBP\_demo.wl". After this, we  have only admissible integrals in the relation. Therefore, if the coefficients of admissible integrals contain linear functions of $\epsilon$, it is safe to set $\epsilon \to 0$. The final relation is thus independent of $\epsilon$.

 In the subsection \ref{subsect:ExampleIBP}, we illustrate how this strategy is applied for finding relevant IBP relations between the admissible integrals of our two-loop HQET example.

\subsection{Integrating out bubbles: cross loop order reduction}
\label{subsection:cross-loop}

%\jmh{Make this subsection more concise}

\begin{figure}[h]    
    \begin{minipage}{0.49\linewidth}
		\centering
		\includegraphics[width=7.4cm,height=3.7cm]{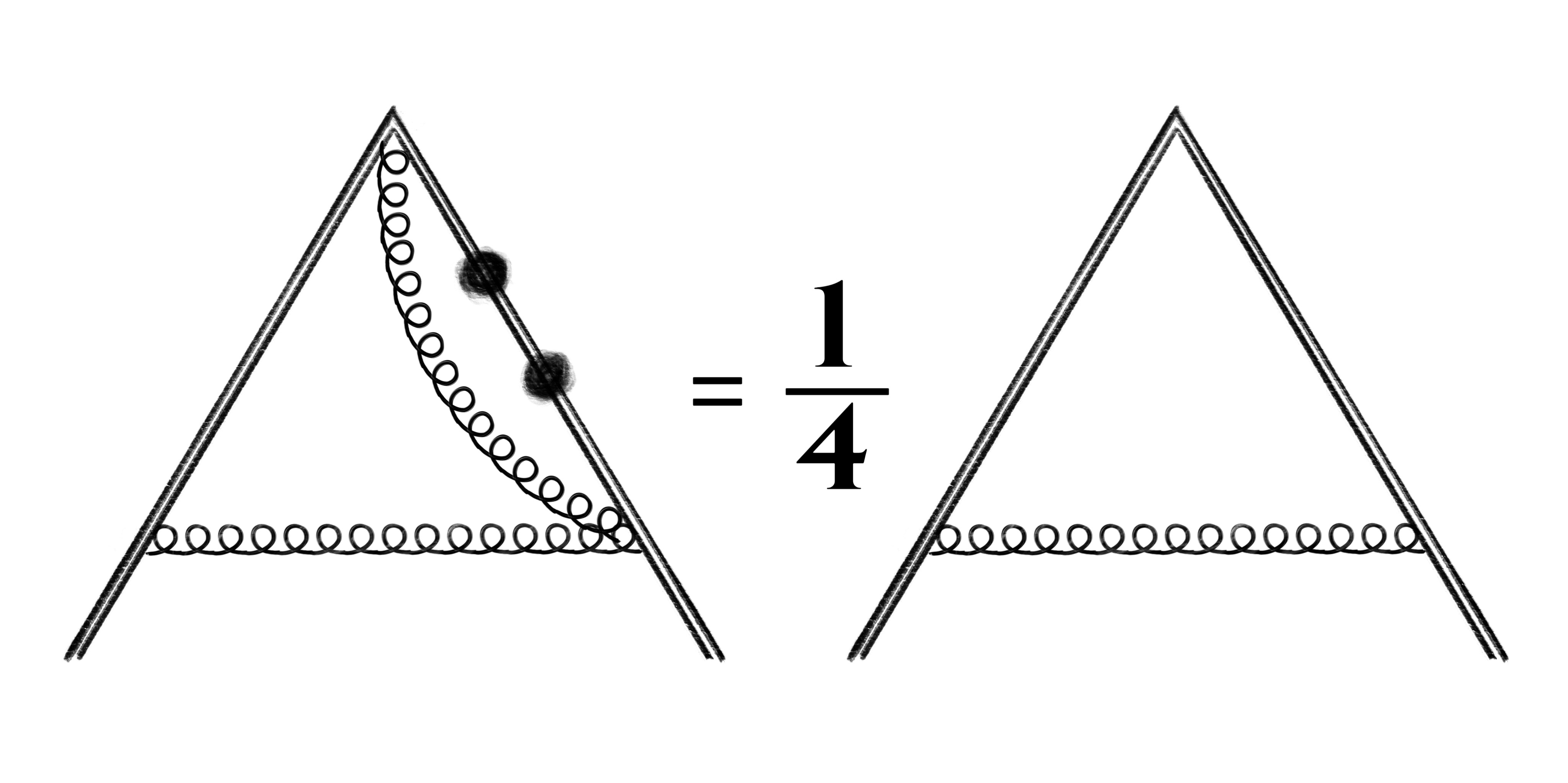} 
		%\caption*{$C_1$}
	\end{minipage}
	\begin{minipage}{0.49\linewidth}
		\centering
		\includegraphics[width=7.4cm,height=3.7cm]{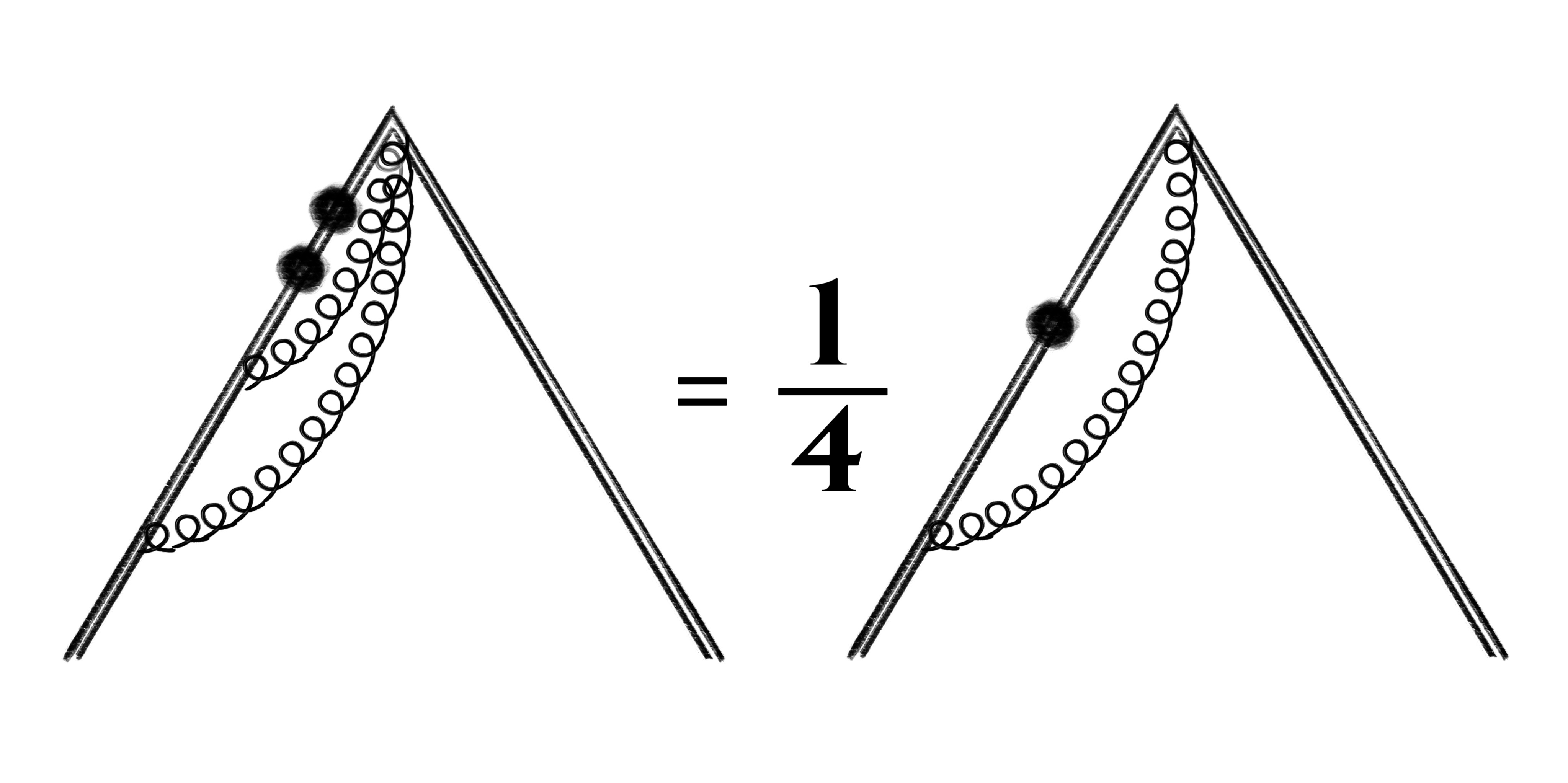}  
		%\caption*{$D_1$}
	\end{minipage}
 \caption{\label{fig:loopreduction} Relations between admissible integrals of different loop orders. 
%\jmh{Add figure for subdiagram relation? The factor of two loops strange.} \ky{might consider replacing "$\rightarrow$" with "$\sim$" and remove the coefficient, which looks more clean}
}
\end{figure}

Another useful technique for generating linear relations for
admissible integrals, is to integrate out bubble sub-diagrams. 
To be specific, for an $L$-loop HQET integral with the loop momenta $k_1$ only appearing in a bubble, with $a_1>2$,
\begin{align}
  \int  \frac{d^{4-2\epsilon} k_1}{i \pi^{2-\epsilon}} \frac{1}{(-2 k_1\cdot v_1+\delta)^{a_1}
  (-(k_1-k_2)^2)} = \frac{1}{(a_1-1)(a_1-2)}\frac{1}{(-2 k_2
  v_1+\delta)^{a_1-2+2\epsilon}},\,.
  \label{eq:D_dim_bubble}
\end{align}
Note that the new propagator's index depends on $\epsilon$ which 
is not always desirable.
%can cause computational difficulties. 
However, for the
  $\epsilon^{-1}$ part only, as we show presently, the following replacement is valid, for $a_1>2$,
\begin{align}
  \int \frac{d^{4-2\epsilon} k_1}{i \pi^{2-\epsilon}}\frac{1 }{(-2 k_1\cdot v_1+\delta)^{a_1}
  (-(k_1-k_2)^2)} \to  \frac{L-1}{L} \frac{1}{(a_1-1)(a_1-2)}\frac{1}{(-2 k_2
  v_1+\delta)^{a_1-2}}\,.
  \label{eq:D_dim_bubble_finite}
\end{align}
The factor $(L-1)/L$ requires a comment. It takes into account the fact that the overall divergence we consider is computed from a volume integral, which in turn depends on the dimensionality, and hence the loop order. The factor compensates for the mismatch when comparing overall divergences of $L$ and $(L-1)$-loop integrals in dimensional regularization.

Applying this relation to the case of our family of crossed ladder integrals, we find the relations 
%\begin{align}
%C_1 = %2 \,
%{\rm T^{(2)}} [1,0,3,0,1,1,0]=\frac{1}{4} G^{(1)}[1,1,1]\,\\
%D_1 = %2 \,
%{\rm T^{(2)}} [1,3,0,0,1,1,0]=\frac{1}{4}G^{(1)}[2,1,0]\,
%\end{align}
shown in Fig.~\ref{fig:loopreduction}.
%\jmh{Also add relation for four-propagator integral?}

\subsection{Application to crossed ladder integrals}
\label{subsect:ExampleIBP}

%Let us now apply this to the crossed ladder integral family.
We construct IBP vectors, or equivalently differential operators as in eq. (\ref{eq:IBP_operator}), which do not double propagators number five and six (see Fig.~\ref{fig:admissiblecrossedladderintegrals} and admissibility condition number \ref{condition2} in subsection \ref{subsec:webs}). 
We find that it suffices to consider those with the lowest scaling dimension in $\{k_1, k_2\}$,  
\begin{align}
    {\bf P_1} &= \partial_{1\mu} k_1^\mu\,, \\
    {\bf P_2} &=   \partial_{1\mu} k_{1\nu} v_1^{[\mu} v_2^{\nu]}\,,\\
    {\bf P_3} &= (\partial_{1\mu} k_{1\mu} + \partial_{2\mu} k_{2\nu}) v_1^{[\mu} v_2^{\nu]}  \,.
\end{align}
%\jmh{Comment on $P_1$. Does it just give trivial relations?}
In the top sector, the IBP relations, 
\begin{align}
{\bf P_{2}} {\rm T^{(2)}}[1,1,1,1,1,1,0]=0 \,,\quad  {\bf P_{3}} {\rm T^{(2)}}[1,1,1,1,1,1,0]=0,    
\end{align}
together with symmetry relations yield
\begin{align}\label{IBPcrossedladder1}
0= & A_2 + B_5
- 2\, B_1 %{\rm T^{(2)}}[1,2,1,0,1,1,0] 
- 2(v_1\cdot v_2)\, B_2 
%{\rm T^{(2)}}[1,1,2,0,1,1,0]
  \,,  \\
0 = & A_3-A_2  \,.
\end{align}
As a result, we keep only one master integral in the six-propagator sector, namely $A_1$, as $A_2$ and $A_3$ can be reduced to lower-sector integrals. 
%\jmh{I introduced the notation $B_5 = {\rm T^{(2)}} [1,2,0,1,1,1,0] $ in our list of admissible integrals.}
It is possible to verify that finite integrals with two or more dots can be reduced in similar ways by applying the same set of IBP vectors on finite seeds which have more dots but fixed scaling dimension.

%\item (1,1,1,0,1,1,0) 

%In this sector we list the admissible integrals with no more than two dots on HQET propagators,
%\begin{align}
%B_1 \,,   \quad B_2 \,, \quad  B_3 = {\rm t^{(2)}}[2,1,2,0,1,1,0,-1,0] \,,  \quad  B_4 =  {\rm t^{(2)}}[2,2,1,0,1,1,0,-1,0] \,. 
%\end{align}
Let us now proceed with the five-propagator sectors.
Using symmetry relations, together with the IBP identities
\begin{align}
&{\bf P_3}  {\rm T^{(2)}}[1,2,1,0,1,1,0]  =0,\;  \quad {\bf P_3} {\rm T^{(2)}}[1,1,2,0,1,1,0]  =0,\; \notag \\
& {\bf P_1} {\rm T^{(2)}}[1,2,0,1,1,1,0] = 0 ,    \;  \quad 
{\bf P_1} {\rm T^{(2)}}[1,3,-1,1,1,1,0]  = 0 , \; \notag \\
&{\bf P_1} {\rm t^{(2)}}[2,2,-1,1,1,1,0,-1,0]  = 0,  \; \quad 
 {\bf P_1} {\rm t^{(2)}}[1,2,-1,2,1,1,0,0,-1] = 0  \,, 
\end{align}
%\ky{a few more added in the $B_5-$ sector} 
we find
\begin{align}
0= & B_4 -  2 (v_1 \cdot v_2) B_1  - B_2   + 2 C_4   \,,   \\  
0= & B_3  + B_1 - 2  C_1  \,, \\
0= &   B_5-2\,C_1 \,, \\
0 = &  2\, B_6 -  C_4 - 3 \, C_2 \,, \\
0 = & B_8 -2\, B_6+ 2\, C_4 \,, \\
0 = &  B_7 - 2\, C_3 \,. 
\end{align}
%\jmh{What is ${\rm T^{(2)}}[1,3,0,0,1,1,0]$? Replace by admissible integrals?} \april{it is $D_1$, which is just an admissible integral.} \ky{Yes.}
This allows us to eliminate $B_{3-8}$ in favor of $B_1, B_2$ and integrals from lower sectors.

Next we move on to four-propagator integrals. 
Due to the presence of one-loop subdiagrams, they can be reduced 
to one-loop integrals, as shown in Fig.~\ref{fig:loopreduction}.

In summary, we find that the admissible crossed ladder integrals can be reduced to the following basis integrals:
\begin{align}
\{ A_1, B_1, B_2 , C_1, C_4  \} \,.
\end{align}
In the next section, we show how to compute these integrals from differential equations.

\section{Algorithm for 
four-dimensional canonical differential equations}
\label{sec:4D_initial}

Having identified the IBP relations between admissible integrals, and hence a set of master integrals,
we can now derive differential equations for the latter.

The HQET integrals we consider depend on a single variable $\phi$.
It is convenient to trade the latter for another variable $x = \exp(i \phi)$
%\jmh{Check sign of exponent.} \ky{(??)}
%scale $x$, namely the exponential of the scattering angle, satisfying 
%\begin{align}
%x+ \frac{1}{x}  =2 \cos \phi  \,. %=  \frac{2 v_1 \cdot v_2 }{ \sqrt{v_1^2 v_2^2}}\,.
%\end{align}
We may differentiate the master integrals with respect to $x$ by means of  
a differential operator acting on the integrand, e.g. 
\begin{align} \label{diffdef}
{\bf d} \equiv \frac{x^2-1}{2} \frac{d}{d x} \equiv  \frac{ (v_1 \cdot v_2 )  v_1^\mu - v_1^2 v_{2}^\mu }{\sqrt{v_1^2 \, v_2^2}} \frac{\partial}{\partial v_{1}^\mu} \,.
\end{align} 
Note that this operator does not double any gluon propagators and preserves the scaling dimension in each loop momenta.  
Thus for any admissible integral $f \in    \mathbb V $,  $ {\bf  d}  f \in  \mathbb V $.  
Given a basis of  master integrals  $\vec f$, one can write down  a linear system of differential equations of degree $ N=  {\rm dim} ( \mathbb V)$ :  
\begin{align}\label{standardDE}
\frac{d}{d x} \vec f = A (x) \vec{f} \,.
\end{align} 
Let us illustrate how this works in the case of the crossed double ladder $A_1$.
The crossed ladder integral family has a basis of five admissible master integrals
\begin{align}
\vec{f} = \{    A_1 \,, B_1 \,, B_2 \,, C_1\,, C_4   \}^{T} \,,
\end{align}
Let us now derive the differential equation that $A_1$ satisfies, and then determine it from now.

Taking the derivative of $A_1$ by means of the differential operator given in \eqref{diffdef}, we have 
\begin{align}
{\bf{d}} A_1 + 2 (v_1 \cdot v_2) A_1 - A_2 - B_5  =0 \,.
\end{align}  
Taking into account eq.(\ref{IBPcrossedladder1}), this becomes
\begin{align}
\frac{d}{d x} A_1 = \frac{2(1+x^2)}{x(1-x^2)} A_1 +  \frac{4}{x^2-1} B_1 + \frac{2 (1+x^2)}{x(x^2-1) }  B_2 \,.
\end{align}
One may iterate this procedure for $B_1, B_2$, and then for $C_1, C_4$.
%Thus our next task is to find differential equations for $B_1, B_2$. We find
%\begin{align}
% \frac{d}{d x} B_1 &= \frac{1+x^2}{x(1-x^2)} B_1 + \frac{2}{x^2-1} B_2 \nn \\
%\frac{d}{d x} B_2 & = \frac{2(1+x^2)}{x(1-x^2)} B_2 + \frac{4}{x^2-1} C_1  \,.  
%\end{align}
% Iterating the above procedure also for $C_1, D_1$, we have all the ingredients needed for eq. (\ref{standardDE}). We find
One finds that the integrals satisfy the differential equations (\ref{standardDE}) with
\begin{align}
    {\mathbf{A}} =   \frac{1+x^2}{1-x^2}
    \begin{pmatrix} 
     \frac{2}{x}  &  -\frac{4}{1+x^2} & -\frac{2}{x} &  0 & 0  \\
    0  &  \frac{1}{x}&  -\frac{2}{1+x^2} & 0  &  0 \\ 
    0  &  0 &   \frac{2}{x}   &   -\frac{4}{1+x^2} &  0   \\
     0 &   0 &  0  &  \frac{1}{ x}  &  -\frac{2}{1+x^2} \\ 
    0  &   0 & 0  & 0   & 0
    \end{pmatrix} 
\,.
\end{align}
In order to solve this system of differential equations, it is useful to first simplify the equations, following \cite{Henn:2013pwa,Caron-Huot:2014lda} (see also \cite{Gehrmann:2014bfa}), or using the {\sc Initial} algorithm \cite{Dlapa:2020cwj}.
%\begin{align} \label{G1Eqs}
%\frac{d}{d x} C_1 &= \frac{1+x^2}{ x(1-x^2)}  B_1 +  \frac{2}{x^2-1}  C_1 \,, \nn \\
%\frac{d}{d x} D_1 & = 0 \,. 
%\end{align}
The system above is of course very simple, so many different methods could be used.
The key idea proposed in \cite{Henn:2013pwa} is to convert the DE to a canonical form.
The latter is obtained by choosing the basis integrals astutely. 
% In fact, if we are primarily interested in one given master integral, say $A_1$, we can
% equivalently consider the (higher-order) homogeneous Picard-Fuchs equation it satisfies.
% The latter can be obtained from (\ref{standardDE}).
% The advantage is that the form of the Picard-Fuchs equation does not depend on the basis choice.
 In fact, in order to find the transformation to a canonical form, under certain assumptions it is sufficient to choose one integral in a good way \cite{Dlapa:2020cwj}.
 Here we can choose
 \begin{align}
 F_1 = \frac12 \left(x- \frac{1}{x} \right)^2 A_1\,.
 \end{align}
 This choice can be justified by an integrand analysis \cite{Henn:2013pwa}, as explained in detail in \cite{Grozin:2015kna}.
 The factor ensures that $F_1$ has constant leading singularities.
 
Applying the {\sc Initial} algorithm, we find that $F_1$ satisfies a fourth order Picard-Fuchs equation,
which defines a rank-four canonical differential system
\begin{align}
\frac{d}{d x} 
    \begin{pmatrix} 
    F_1\\ F_2 \\ F_3 \\F_4 
    \end{pmatrix} =    
    \begin{pmatrix} 
 0     & \frac{2}{x}   & 0                           &  0 \\
  0    &       0        & \frac{1+x^2}{x(1-x)(1+x)}   & 0  \\ 
   0   &      0         &                  0           &  \frac{4}{x} \\
    0  &       0        &            0                 &    0
    \end{pmatrix} 
    \begin{pmatrix} 
    F_1\\ F_2 \\ F_3 \\F_4 
    \end{pmatrix} \,, \label{eq:canonicalcrossedladder}
    \end{align}
    where 
   \begin{align}
    \{ F_1, F_2, F_3, F_4 \} = \left\{ \frac{(x^2-1)^2}{2\, x^2} A_1,  \frac{x^2-1}{ x} \Big( B_1 + \frac{1+x^2}{2 \, x} B_2 \Big)  , \frac{2(x^2-1)}{ x} C_1,  C_4  \right\}\,.
\end{align}
We see that eq. (\ref{eq:canonicalcrossedladder}) is in canonical form. Moreover, the matrix is block triangular \cite{Caron-Huot:2014lda}.
This means that we can iteratively solve this equation.
At each integration, there is one boundary constant to be fixed.
The first integral, $F_4$, is constant, and we find its value by direct computation. The other boundary constants are fixed by observing that $F_1, F_2, F_3$ vanish at $x=1$.
In this way, we find
\begin{align}
F_1 = \frac{1}{2} \left[  \frac13 \ln^3 x +  \ln x\, {\rm Li}_2 (x^2) -  {\rm Li}_3 (x^2) +  \zeta_2 \ln x +  \zeta_3 \right] \,.
\end{align} 
This is in agreement with the known answer for $A_1$, see e.g.  eq. (27) in \cite{Correa:2012nk}.   %\jmh{Provide reference with explicit formula.}
%\jmh{The standard form for this integral is written with $1-x^2$ arguments of the polylogarithms. That's preferable, because otherwise one might think there is a branch cut at $x=1$.} \ky{This expression agrees with the formula in the reference \cite{Correa:2012nk}.}

%\jmh{Let's first solve for $A_1$. We could then comment on solving for $B_2$ later, if we want.}
This determines four of the five admissible master integrals.
In case one wished to solve for the remaining one, one can do so by repeating the above steps with
 $F_5 = \frac{1}{2} (x- 1/x)^2 B_2 $ as the input to the {\sc Initial} algorithm.
 In this case, one finds
%In addition, $F_5 = \frac{1}{4} (x- 1/x)^2 B_2 $ is also an admissible UT integral, it couples to $F_{3}$ and $F_4$ forming a rank-3 system,
\begin{align}
  \frac{d}{d x}  \begin{pmatrix} 
    F_5\\ F_3 \\ F_4
    \end{pmatrix} =    
    \begin{pmatrix} 
  0    & \frac{1}{x}   &  0             \\
   0   &         0      &    \frac{4}{x} \\
    0  &      0          &       0
    \end{pmatrix}
    \begin{pmatrix} 
    F_5 \\ F_3 \\ F_4  
    \end{pmatrix}   \,.
\end{align}
Solving this system, we find
\begin{align}
 F_5  =     \frac{1}{4} \ln^2 x\,.
 \end{align}
 %\jmh{Check that this agrees with Appendix A.2 of \cite{Correa:2012nk}.} \ky{where can I find the result for B2? }
Hence we have found the complete basis of admissible UT integrals in the crossed ladder family.

 \section{Three-loop application}
 \label{sec:3loop}

Here we apply our new four-dimensional method to three-loop integrals initially computed in \cite{Grozin:2014hna,Grozin:2015kna}.
We find that in comparison, much fewer master integrals are needed,
and the computation requires only a fraction of the computing
time. For example, the standard differential equation approach for our three-loop HQET example via a publicly-available multiple-thread IBP solver took several hours, while our approach only took minutes with one core on the same machine.

\begin{figure}[H] 
    \centering 
    \includegraphics[width=5cm,height=5cm]{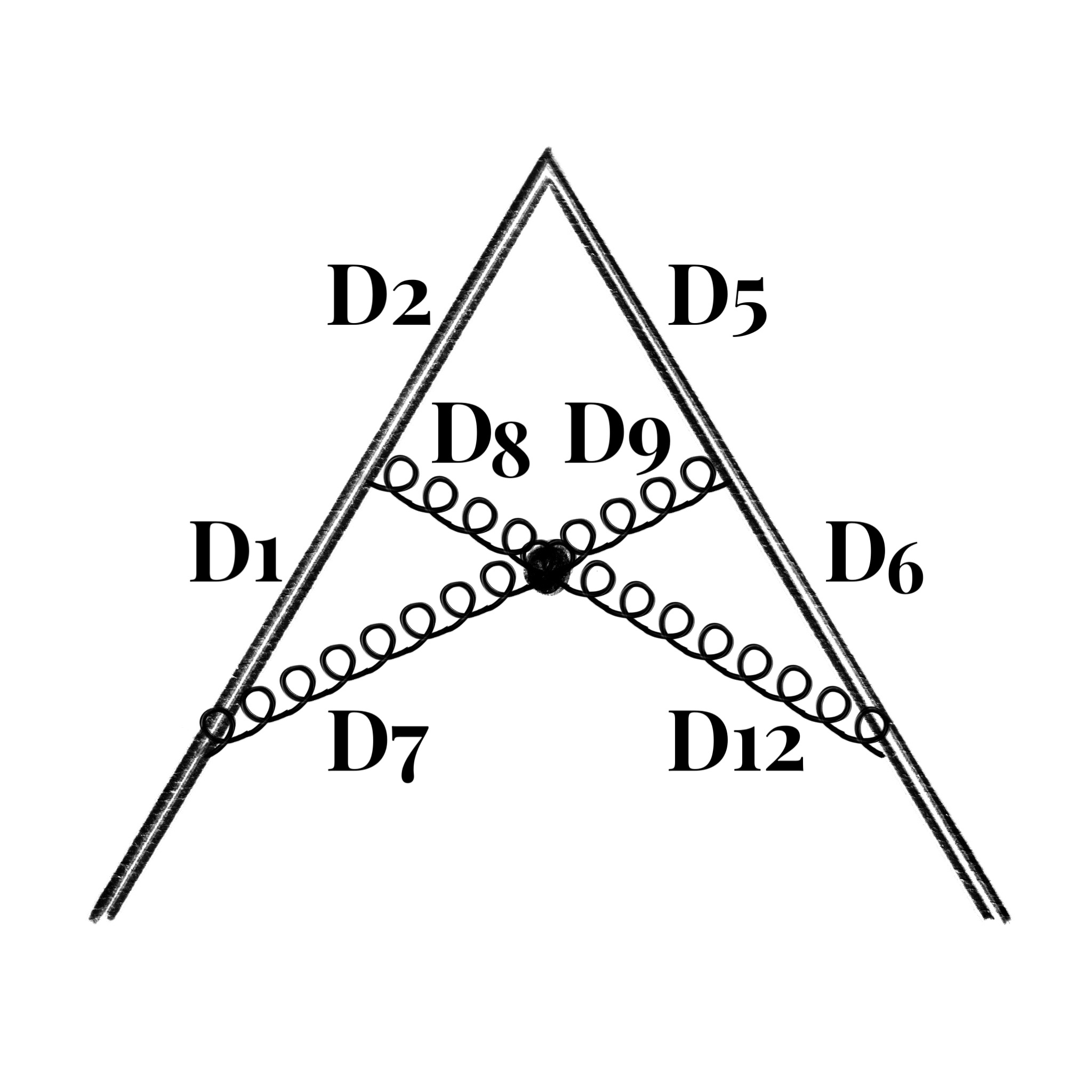} 
\caption{Three-loop HQET integral example}
\label{3l_example}
\end{figure}

The propagators of the three-loop integrals in \cite{Grozin:2015kna} are,
\begin{gather}
  \label{eq:1}
  D_1= -2 k_1 v_1+\delta\,,\quad D_2=-2 k_2 v_1+\delta\,,\quad D_3 =-2
  k_3 v_1+\delta\,,              \nonumber \\
D_4= -2 k_1 v_2+\delta\,,\quad D_5=-2 k_2 v_2+\delta\,,\quad D_6 =-2
  k_3 v_2+\delta\,,              \nonumber \\
D_7= -k_1^2\,,\quad D_8=-(k_1-k_2)^2\,,\quad  D_9=-(k_2-k_3)^2\,,
\nonumber \\
D_{10}= -(k_1-k_3)^2\,,\quad D_{11}=-k_2^2\,,\quad  D_{12}=-k_3^2 \,,  
\label{eq:3loop_propagator}
\end{gather}
and the integral family is defined via
\begin{gather}
  \label{eq:2}
  G[a_1 ,\ldots , a_{12}]= \int \frac{d^{4-2\epsilon}
    k_1}{i\pi^{2-\epsilon}}\frac{d^{4-2\epsilon}
    k_2}{i\pi^{2-\epsilon}}\frac{d^{4-2\epsilon}
    k_3}{i\pi^{2-\epsilon}} \frac{1}{D_1 ^{a_{1}} \ldots D_{12}^{a_{12}}}\,.
\end{gather}
For example, say, our goal is to calculate the $\epsilon^{-1}$ part of
$G[1,1,0,0,1,1,1,1,1,0,0,1]$, which is an admissible integral (see Fig.~\ref{3l_example}). To set
up a four-dimensional differential equation system, we first list four-dimensions IBPs for
admissible integrals, as described in section
\ref{sec:integral_relation}, with a bound on the propagator
indices. Besides the four-dimensional IBP relations, we also use the cross loop order
relations introduced in the section \ref{sec:integral_relation}. We find that some specific integrals
in two sectors %$(1, 0, 0, 0, 1, 1, 0, 1, 1, 0, 0, 1)$ and
               %$(1,0,0,0,1,1,1,0,1,1,0,0)$
 can be reduced to lower loop integrals.

Combining the four-dimensional IBPs and cross loop-order relations, we get a linear
system for  the $\epsilon^{-1}$ part of the admissible integrals. This
linear system is very sparse, free of the parameter $\epsilon$, and
can be solved by {\sc FiniteFlow}
\cite{Peraro:2016wsq,Peraro:2019svx} in two minutes with one CPU
core. We find $13$ irreducible integrals (cf. the auxiliary file {``\tt{output/3lHQET\_4D\_MI.txt}"}) that satisfy the differential equation,
%\begin{gather}
  %\{I_{1},\ldots , I_{13}\}=\bigg\{  g[2, 2, 0, 0, 1, 1, 1, 1, 1, 0, 0, 1, -1, 0, 0],\quad  G[1, 1, -1, 0, 1, 2, 
%1, 1, 1, 0, 0, 1], \nonumber \\
%G[1, 1, 0, 0, 1, 1, 1, 1, 1, 0, 0, 1],\quad g[3, 1, 0, 0, 2, 2, 0, 1, 1, 0, 0, 1, 0, 0, -1],\nonumber \\ G[3, 1, -1, 
%0, 1, 2, 0, 1, 1, 0, 0, 1], \quad g[1, 0, 0, 0, 3, 2, 1, 1, 1, 0, 
%0, 1, 0, 0, -1],\nonumber\\ G[1, 0, -1, 0, 2, 2, 1, 1, 1, 0, 0, 1], 
%G[1, -1, 0, 0, 3, 1, 1, 1, 1, 0, 0, 1], G[1, 0, 0, 0, 2, 1, 
%1, 1, 1, 0, 0, 1] ,\nonumber\\  H^{(2)}[2, 0, 1, 1, 1, 0, 1],\quad H^{(2)}[1, 0, 2, 1, 1, 
%0, 1], \quad \tilde{H}^{(1)}[1, 1, 1], \quad \tilde{H}^{(1)}[0, 2, 1] \bigg\}
%\label{eq:3loop_basis}
%\end{gather}
\begin{equation}
    \frac{d}{dx} \vec I =A(x) \vec I    \,,
\end{equation}
where $A$ is a $13\times 13$ matrix. 
%which is independent of $\epsilon$. 
The explicit expression of $A$ is given in the auxiliary file {``\tt{output/3lHQET\_4D\_DE.txt}"}. 

A straightforward integrand analysis, as explained in detail in \cite{Grozin:2015kna},
suggests that the integral 
\begin{equation}
J_1 \equiv \frac{(-1 + x)(1 + x)}{x}\ \lim_{\epsilon\to 0} \epsilon \, G[1, 1, 0, 0, 1, 1, 1, 1, 1, 0,
0, 1] \,,
\end{equation}
 is UT and has constant leading singularity.
This knowledge allows us to apply
 the INITIAL algorithm and find a basis of only $8$ UT integrals (in  {``\tt{output/3lHQET\_4D\_UT.txt}"}). 
 %which are given explicitly in {\tt{``output/UT06-17-22.txt"}}. 
 They
 satisfy a  simple canonical differential equations  
$\frac{d}{dx} \vec J =B \vec J$, 
%\begin{align}
%\label{example:3lHQET_DE}
 %  B &\equiv 
  % \begin{pmatrix}
 %  0 & -\frac{2}{(-1+x)(1+x)} &0 & 0& 0& 0&0 & 0 \\
   % &                          & \frac1x   & 0 & 0 & 0 & 0 & 0 \\
   % &                          &     &  \frac1x & \frac{2x}{(-1+x)(1+x)} & 0 & 0 & 0 \\ 
   % &                          &     &    &      &  \frac1x &  0 &  0\\ 
   % &                          &   &  &  &       &  \frac1x & 0 \\
   % &                          &   &  &  &        &  & \frac1x \\
   % &                          &   &  &  &        &    &  - \frac1x  \\
    % &                          &   &  &  &        &    &   
  % \end{pmatrix}\, ,
%\end{align}
where $B$ is an extremely sparse up-triangular matrix, 
%(with each empty entry as $0$),
\begin{align}
\label{example:3lHQET_DE}
   B &\equiv 
   \begin{pmatrix}
   0 & -\frac{2}{(-1+x)(1+x)} &0 & 0& 0& 0&0 & 0 \\
  0  &           0               & \frac1x   & 0 & 0 & 0 & 0 & 0 \\
   0 &            0              &   0  &  \frac1x & \frac{2x}{(-1+x)(1+x)} & 0 & 0 & 0 \\ 
   0 &             0             &  0   &  0  &   0   &  \frac1x &  0 &  0\\ 
   0 &              0            &  0 & 0 & 0 &    0   &  \frac1x & 0 \\
   0 &               0           &  0 & 0 & 0 &     0   & 0 & \frac1x \\
   0 &                0          &  0 & 0 & 0 &      0  &  0  &  - \frac1x  \\
   0  &                0          & 0  & 0 & 0 &      0  &  0  &  0 
   \end{pmatrix}\, ,
\end{align}

Solving the differential equations and fixing the boundary constants,
we find that there is  degeneracy%in the basis $\vec{J}$, in
                                %particular $J_7 = -J_6 =  \frac{8}{3}
                                %\ln x, J_5 = -J_4  = \frac{4}{3}
                                %\ln^2 x  $, 
, and hence the dimension of the finite system is reduced to six.  The solution for the top integral $J_1$ can be easily found as,
\begin{align}
    J_1 &= \frac83 H_{-1, -2, 0, 0} - \frac83  H_{-1, 2, 0, 0} 
    + \frac83 H_{1, -2, 0, 0}  - \frac83 H_{1, 2, 0, 0} 
  - \frac83 H_{-1, 0, 0, 0, 0} - \frac83 H_{1, 0, 0, 0, 0}   \notag\\ 
  & - \zeta_4 \,  \ln (1 - x) + \zeta_4 \, \ln(1 + x)
  + \frac43 \zeta_3\, \ln( 1 - x) \ln x 
  - \frac43 \zeta_3 \, \ln x  \ln(1 + x) \notag \\
  &+ \frac83  \zeta_3\,  {\rm Li}_2 (x)  
  - \frac23 \zeta_3 \, {\rm Li}_2 (x^2)\,.
\end{align}
where $H_{a_1, \ldots, a_n} \equiv {\rm HPL} [\{a_1, \ldots, a_n\}, x]$ are the standard harmonic polylogarithms (HPLs). This analytic result is consistent with that in \cite{Grozin:2015kna}.
%\jmh{The explicit integral in the ancillary files is $f_{64}$.}
%\jmh{Has the agreement been checked numerically?}
%\april{Yes, $J_1$ agree with $f_{64}$  numerically.}

It is interesting to compare our approach with that in the ref.~\cite{Grozin:2015kna}. In ref.~\cite{Grozin:2015kna}, in order to compute
$J_1$, $39$ master integrals are needed.
%\jmh{From looking at the ancillary files of \cite{Grozin:2015kna}, and in particular the matrix structure, it seems to me that at most $39$ integrals would be needed to compute $J_1$ there. Can you please double check?} %\ky{Yes.}
%Further efforts are
%needed to convert all those
%integrals  to the UT form and to find the boundary
%conditions. 
Here, in our approach, to get the $\epsilon^{-1}$ order of
$J_1$, eventually we just need $6$ (or $13$) master integrals. 
%It clearly demonstrates the efficiency and simplicity of our method. 

 \section{Summary and outlook}

%In this paper, we invent a new canonical differential equation method for evaluating the divergent part of Feynman integrals. 
%More specifically, we focus on Feynman integrals with the overall divergence only, which is natural for the eikonal web diagrams. 
%Using the cutting-edge IBP techniques with the graded IBP operators and the syzygy, we develop a method to find linear relation of the divergent part of these integrals. 
%Furthermore we carry out a reduction on the loop order so that the linear relations would be further simplified. 
%These linear relations are about the $\epsilon^{-1}$ part of the integrals and thus independent of $\epsilon$. So we obtain a differential equation without the parameter $\epsilon$.
%Then we develop a new {\sc initial} algorithm to convert the differential equation to canonical differential equations. 
%The resulting differential equation matrix would be up triangular and have all diagonal elements zero. This structure guarantees that the divergent part of these integrals can be solved by iterated integrals. 

\begin{figure}[H] 
    \centering 
    \includegraphics[width=0.9\textwidth]{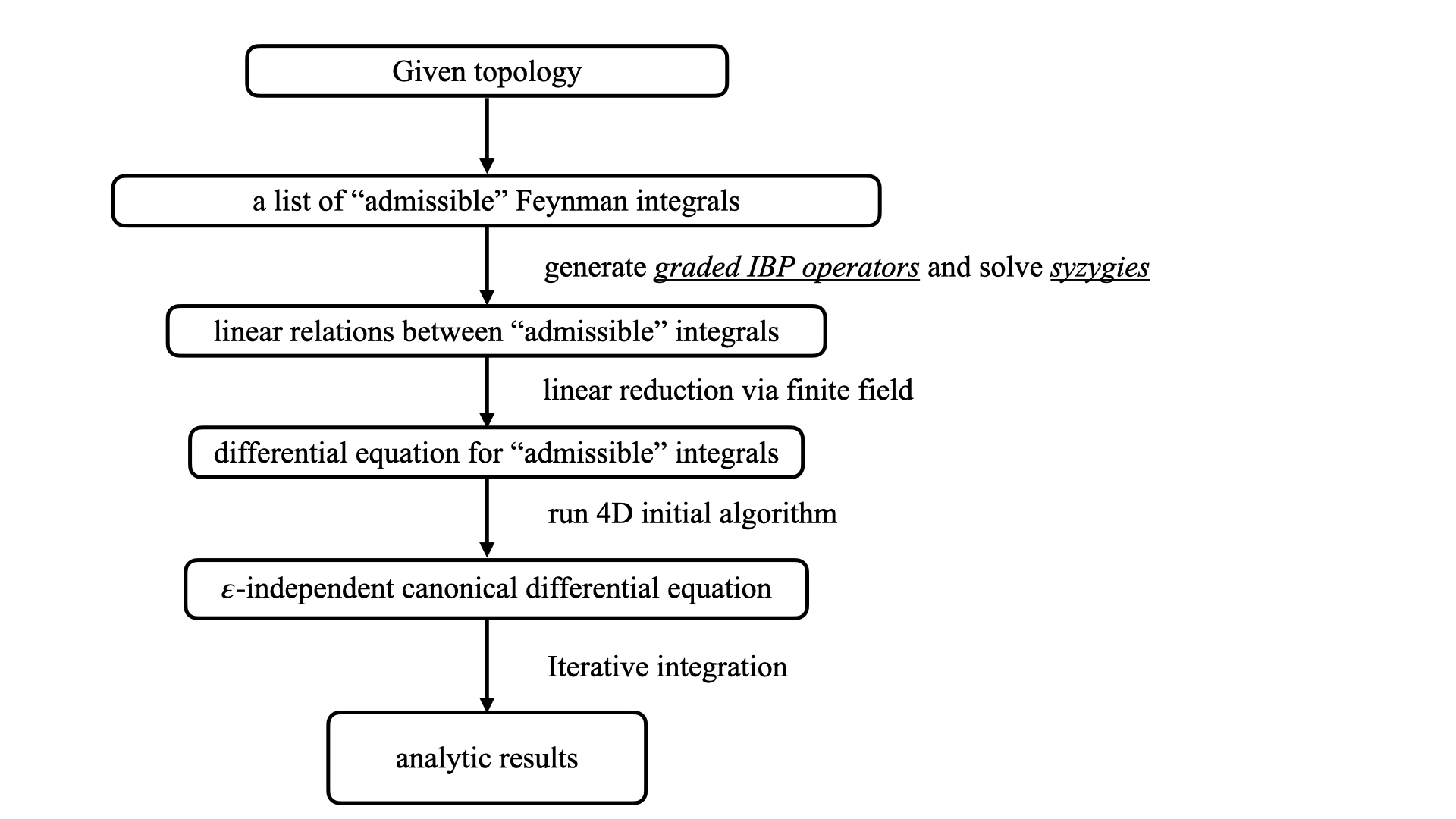} 
\caption{The algorithm flow chart of our method}
\label{flowchart}
\end{figure}

We presented a method for computing the leading divergent part of Feynman integrals that are free of subdivergences. 
Our method leverages simplifications that occur in limit as the dimensional regulator is taken to zero.
We effectively use four-dimensional IBP relations and differential equations.
This leads to substantial improvements: fewer master integrals are needed, and the IBP relations are faster to solve.
The presented pedagogical examples and reproduced three-loop HQET integrals from the literature are proofs of concept. This method achieves the computation by the steps in Figure \ref{flowchart}.
%Our moethod is efficient for generating and solving IBP relations, and powerful to find integrals whose $\epsilon^{-1}$ part has the uniform transcendental weight. 
%Hence we believe that this method will play an important role for the  applications like computing HQET integrals of the higher loop orders, the integrals appearing in the effective theory of gravitational waves. 
%We have the following outlook for the future development of our method:
%The following promising extensions of the work in this paper.
We find the following extensions of the work in this paper promising:
\label{sec:summary}
\begin{enumerate}
\item Apply the method to cutting-edge applications that go beyond the state of the art. Potential applications include
anomalous dimensions of composite operators, and the soft anomalous dimension matrix \cite{Liu:2022elt}.
\item Extend the method to Feynman integrals with subdivergences, truncating the Laurent expansion in the dimensional regulator at a given order.
This requires a careful identification of the relevant regions of loop integration. This will likely lead to an increase in the number of finite master integrals to be considered, but we still expect a significant simplification compared to the general, $D$-dimensional case.
\item Combine the method with insights into the structure of the Feynman graph polynomials from tropical geometry. The latter allow to identify the relevant divergent regions and to write their (finite) coefficients in terms of integrals that appear naturally from the geometry \cite{Schultka:2018nrs,Arkani-Hamed:2022cqe}.
\end{enumerate}

\section*{Acknowledgments}
We acknowledge Christoph Dlapa, David Kosower, Zhao Li, Zhengwen Liu,  Xiaoran Zhao, and Yu Wu for enlightening discussions. This research received funding from the European Research Council (ERC) under the European Union's Horizon 2020 research and innovation programme (grant agreement No 725110), {\it Novel structures in scattering amplitudes}.  YZ is supported from the NSF of China through Grant No. 11947301, 12047502, 12075234, and the Key Research Program of the Chinese Academy of Sciences, Grant No. XDPB15. YZ also acknowledges the Institute of Theoretical Physics, Chinese Academy
of Sciences, for the hospitality through the ``Peng Huanwu visiting professor program”.

\bibliographystyle{JHEP}
\bibliography{refs}
\end{document}